\documentclass[useAMS,usenatbib]{mn2e}

\usepackage{natbib}
\usepackage{epsfig}
\usepackage{calc}
\usepackage{amssymb}
\usepackage{amstext}
\usepackage{amsmath}
\usepackage{multicol}
\usepackage{pslatex}
\usepackage{fancyhdr}
\usepackage{longtable}
\usepackage{pdflscape}
\usepackage{graphics}
\usepackage{graphicx}

\begin{document}
\title{
{STEREO observations of long period variables}
}

\author[K. T. Wraight, D. Bewsher, Glenn J. White, W. Nowotny, A. J. Norton and C. Paladini]{K. T. Wraight$^{1}$\thanks{e-mail:
k.t.wraight@open.ac.uk}, D. Bewsher$^{2}$, Glenn J. White$^{1,3}$, W. Nowotny$^{4}$, A. J. Norton$^{1}$  and \newauthor C. Paladini$^{4}$\\
$^{1}$Department of Physical Sciences, The Open University, Milton Keynes, MK7 6AA, UK\\
$^{2}$Jeremiah Horrocks Institute, University of Central Lancashire, Preston, Lancashire, PR1 2HE, UK\\
$^{3}$Space Science and Technology Department, STFC Rutherford Appleton Laboratory, Chilton, Didcot, Oxfordshire, OX11 0QX, UK\\
$^{4}$Department of Astronomy, University of Vienna, T\"urkenschanzstrasse 17, 1180 Vienna, Austria}

\pagenumbering{arabic}
\maketitle

\begin{abstract}
Observations from the Heliospheric Imagers (HI-1) on both the \textit{STEREO} spacecraft have been analysed to search for very long period large amplitude stellar variability, finding 6 new candidates.  A total of 85 objects, mostly previously known Mira variables, were found to show convincing variability on time scales of over a hundred days.  These objects range in peak brightness from about 4th magnitude to 10th magnitude in \textit{R} and have periods between about 170~days and 490~days.  There is a period gap between 200 and 300~days where no objects were found and this is discussed.  15 of the Miras in the sample are previously recorded as having variable periods and the possibility for these and 2 other stars to have undergone a period change or to be irregular is discussed.  In addition to the 6 stars in the sample not previously recorded as variable, another 7 are recorded as variable but with no classification.  Our period determination is the first to be made for 19 of these 85 stars.  The sample represents a set of very long period variables that would be challenging to monitor from the Earth, or even from Earth orbit, owing to their position on the Ecliptic Plane and that their periods are often close to a year or an integer fraction thereof.  The possibility for the new candidates to possess circumstellar shells is discussed.
\end{abstract}  

\begin{keywords}
techniques: photometric  -- catalogues -- stars: oscillations -- stars: late-type
\end{keywords}

\sloppy

\section{\uppercase{Introduction}}
\label{sec:introduction}

\noindent
Roughly four hundred years ago the first variable star was recognised 
due to its changing brightness \citep{hoffleit1997}. This red object, showing 
striking light variations in the visual, was later named $o$~Cet and 
became the first member of the variability class called long period 
variables (LPVs) today. Being easily detectable because of the large 
photometric amplitudes they were intensively studied thereafter and 
represent now a prominent group within the General Catalogue of Variable 
Stars (Samus et al. 2012). It is known nowadays that LPVs are stars of 
low to intermediate main sequence mass ($\approx$0.8--8\,$M_{\odot}$) in 
a quite late stage of stellar evolution, the Asymptotic Giant Branch 
(AGB). Such late-type giants populate regions of high luminosities 
(several 10$^3\,L_{\odot}$) and low effective temperatures (below 
$\approx$4000\,K) in the Hertzsprung-Russell diagram. During the AGB 
evolution, the stars start to pulsate which causes the pronounced 
photometric variations \citep[e.g.,][]{olivier2005}. Conventionally, 
LPVs were subclassified into Miras, semiregular variables, and irregular 
variables with the visual amplitude and the regularity of the lightcurve 
as criteria. An important step forward in our understanding of LPVs was 
provided by the large surveys of the Magellanic Clouds and the Galactic 
Bulge \citep[e.g.,][]{wood2000,ita2004,groenewegen2005,matsunaga2005}, recently updated with space-based IR data \citep{riebel2010}. From these we know that AGB variables constitute a 
few period-luminosity relations. The most prominent LPVs, namely the 
Mira variables, can be found along the sequence for fundamental mode 
pulsators. The majority show large-amplitude variations ($\Delta V$ of 
a few magnitudes, $\Delta K$~$\approx$~0.4\,--\,1$^{\rm mag}$) on 
time-scales of 100\,--\,600 days \citep[e.g.,][]{whitelock2000}. The 
$P$-$L$ relation for Miras was determined rather precisely based on a 
number of well-characterised LMC objects by \citet{whitelock2008} but 
could be investigated also in other Local Group Galaxies \citep[e.g.,][]{lorenz2011} and even in systems as distant as Cen\,A \citep{rejkuba2004}. 
Apart from pronounced mass loss \citep[e.g.,][]{nowotny2011} AGB stars 
are also characterised by the occurence of He-shell flashes \citep[thermal pulses;][]{lattanzio2004}. Not only are these responsible for 
changing the atmospheric chemistry from O-rich to C-rich via dredge-up 
of carbon from the interior, thermal pulses are also suspected to be the 
reason for period changes in observed lightcurves \citep[e.g.,][]{templeton2005,uttenthaler2011,lebzelter2011b}.

The data used in this study comes from NASA's \textit{STEREO} mission, which aims to image the Sun's corona in 3D and observe Coronal Mass Ejections (CMEs) from the surface of the Sun out to the Earth's orbit \citep{kaiser2008stereo}.  The two satellites are in different heliocentric orbits, one slightly inside the Earth's orbit (\textit{STEREO}-Ahead) and one slightly outside (\textit{STEREO}-Behind) and the angle between each satellite, the Sun and the Earth increases by about 22.5 degrees every year.  Photometry of background stars in the images is possible as the calibration of the Heliospheric Imager cameras has been performed to a very high standard \citep{brown2009,bewsher2010,bewsher2012,wraight2011,wraight2012}.  Having two nearly identical satellites in different heliocentric orbits provides greater phase coverage of long period variables, with greater homogeneity than achievable from observations conducted from the ground.

In the following sections we outline the characteristics of the \textit{STEREO}/HI-1 observations and explain how the sample was extracted and analysed.  We provide periods for 85 stars showing large amplitude variability ($\ge 0.3$ magnitudes) with periods longer than 100~days.  The majority of these are known to be Miras or semi-regular variables.  We estimate times of maximum brightness, where visible.  Our analysis of patterns found in the sample focuses on the errors obtained for the periods and also on a gap in the periods found between 200 and 300~days.  We discuss a number of individual stars, in particular those which are known to have varying periods, those which have not previously been classified and those found here to be variable for the first time.

\section{\uppercase{Observations and Analysis}}
\label{sec:method}

\subsection{Characteristics of \textit{STEREO}/HI-1 observations}

\noindent
The Heliospheric Imagers are described in detail in \citet{eyles2009} but we summarise the main features here for convenience.  The field of view of the \textit{STEREO}/HI-1 cameras is 20 degrees by 20 degrees, centred 14 degrees away from the Sun's centre.  The aperture of the cameras is just 16~mm and the focal length is 78~mm.  Each CCD is 2048x2048 pixels, which provides a plate scale of 35 arc-seconds per pixel.  The images are binned 2x2 on-board, to reduce the bandwidth, and as a result the final images received have a resolution of 70 arc-seconds per pixel.  Each image is the result of 30 summed exposures, with each exposure lasting 40 seconds, and one image is produced every 40 minutes.  In this way, stars as faint as 12th magnitude in \textit{R} can be observed but only the very brightest stars will saturate the CCDs.  Stars of about 4th magnitude may show some systematic effects due to saturation, whilst 3rd magnitude stars often show systematic effects but may be usable whereas stars of 2nd magnitude and brighter are unusable.  We use the \textsc{NOMAD1} catalogue \citep{zacharias2004} to determine which stars to observe, selecting all those listed as 12th magnitude or brighter in the \textit{R} band within our field of view.  Aperture photometry is performed on each star as described in \citet{bewsher2010} and the data reduction pipeline is summarised in \citet{wraight2011}.  This provides an extremely useful resource, with photometry of almost 900,000 stars along the Ecliptic Plane.

\begin{figure}
\resizebox{8cm}{!}{\includegraphics{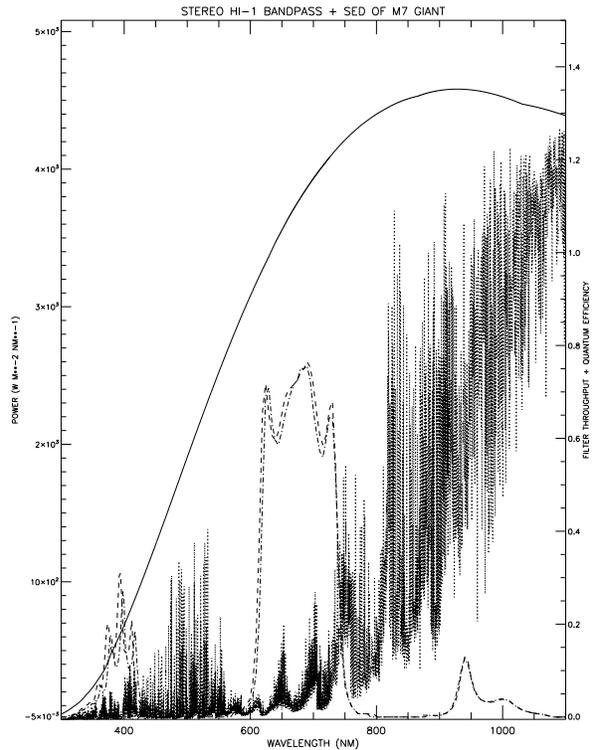}}
\caption{Plot of the filter throughput convolved with the quantum efficiency of the CCDs for \textit{STEREO}/HI-1A, shown in a dashed line, (\textit{STEREO}/HI-1B is almost identical \protect\citep{bewsher2010}, shown here in a dot-dashed line).  This is set against a synthetic spectral energy distribution of a red giant with MK spectral type M7 \protect\citep{fluks1994}, shown in a dotted line, with the continuum emission in a solid line.}
\label{fig1}
\end{figure}

The \textit{STEREO}/HI-1 imagers have an unusual spectral bandpass (Figure \ref{fig1}).  They are most sensitive inbetween 630~nm and 730~nm but also have some sensitivity in the blue, around 400~nm, and in the infra-red at around 950~nm.  The sensitivity in the blue is very useful for observing hot stars \citep{wraight2012} but it is the sensitivity in the infra-red that makes the observations of many of the very cool stars in this paper possible.  Some of the stars in our sample are cool enough, or sufficiently obscured by circumstellar material, to be fainter than 12th magnitude in the \textit{R} band and do not feature in the \textit{STEREO} database directly.  In these cases the large pixel size of the \textit{STEREO}/HI-1 imagers, combined with the sensitivity near 950~nm, allows them to be observed indirectly through blending with a nearby star that is in the database.  As Mira variables are known to change their temperature and spectral type, the magnitude of the variability observed by \textit{STEREO}/HI-1 is not reliable.  This can be seen by referring to synthetic spectra of red giants \citep{fluks1994}.  There is a significant difference in the emission in the region 630~nm to 730~nm between the spectral type M6, where there is still some emission, and M7 and later, where there is relatively little emission.  A star varying across this threshold will therefore have an exaggerated amplitude.  Giants of spectral type M7 or later are seen almost entirely through their emission in the infra-red and may show different variability to that seen in other parts of the spectrum, whilst giants of M6 and earlier may show variability due to processes affecting both the red and the infra-red.  In particular, it is worth noting that some of the faintest stars in our sample, including many of the new candidates, are so faint that only deep all-sky surveys e.g. \citet{zacharias2004} and \citet{monet2003} have detected them in visible light.  Even in the infra-red observations are scarce, mostly from the \textit{IRAS} mission \citep{neugebauer1984}, the \textit{AKARI} mission \citep{murakami2007} and in the 2-Micron All-Sky Survey (2MASS) \citep{skrutskie2006}.

\subsection{Quality of \textit{STEREO}/HI-1 data}

In order to determine a period for the candidate variables, it was necessary to use undetrended data and the only processing that was applied was the exclusion of obvious outliers and systematics, where possible.  This is because the polynomial detrending normally used to clean the lightcurves removes long period variability.  As a result, many of the lightcurves show indications of artificial trends, often a result of the flat-fielding breaking down near the edges of the detectors.  Other sources of noise affect the long period signals being searched for to a lesser degree, most notably de-pointing events associated with micrometeorite hits, which are more common in the data from the \textit{STEREO}-Behind satellite, \textit{STEREO}/HI-1B, also observed by \citet{davis2012}.  The presence of artificial trends and the ability to recognise them as such also limits the number of maxima that can be reliably observed for the stars in the sample and might also influence the times of maximum light, in those cases where the effect is small enough to be reasonably sure a maximum is genuinely being observed.

\subsection{Data analysis}
\label{sec:analysis}

\begin{figure}
\resizebox{8cm}{!}{\includegraphics{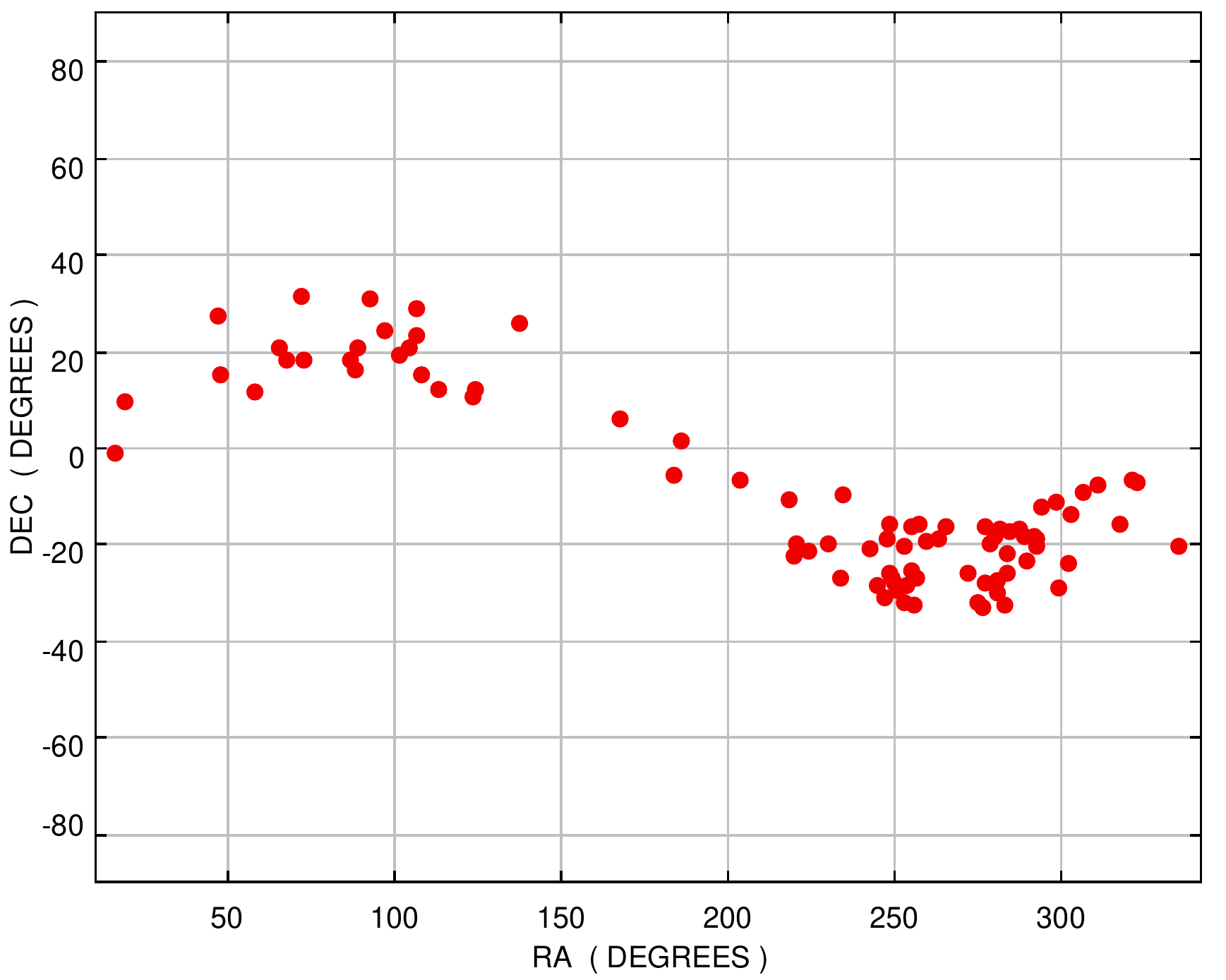}}
\caption{Plot of the locations of the 85 stars in the sample on the sky, with Right Ascension and Declination given in degrees.  As \textit{STEREO}/HI-1 only observes stars within 10 degrees of the Ecliptic Plane, this is where the candidates are found.  Note that more are observed near to the Galactic Centre.}
\label{fig2}
\end{figure}

\begin{figure*}
\resizebox{5cm}{!}{\includegraphics{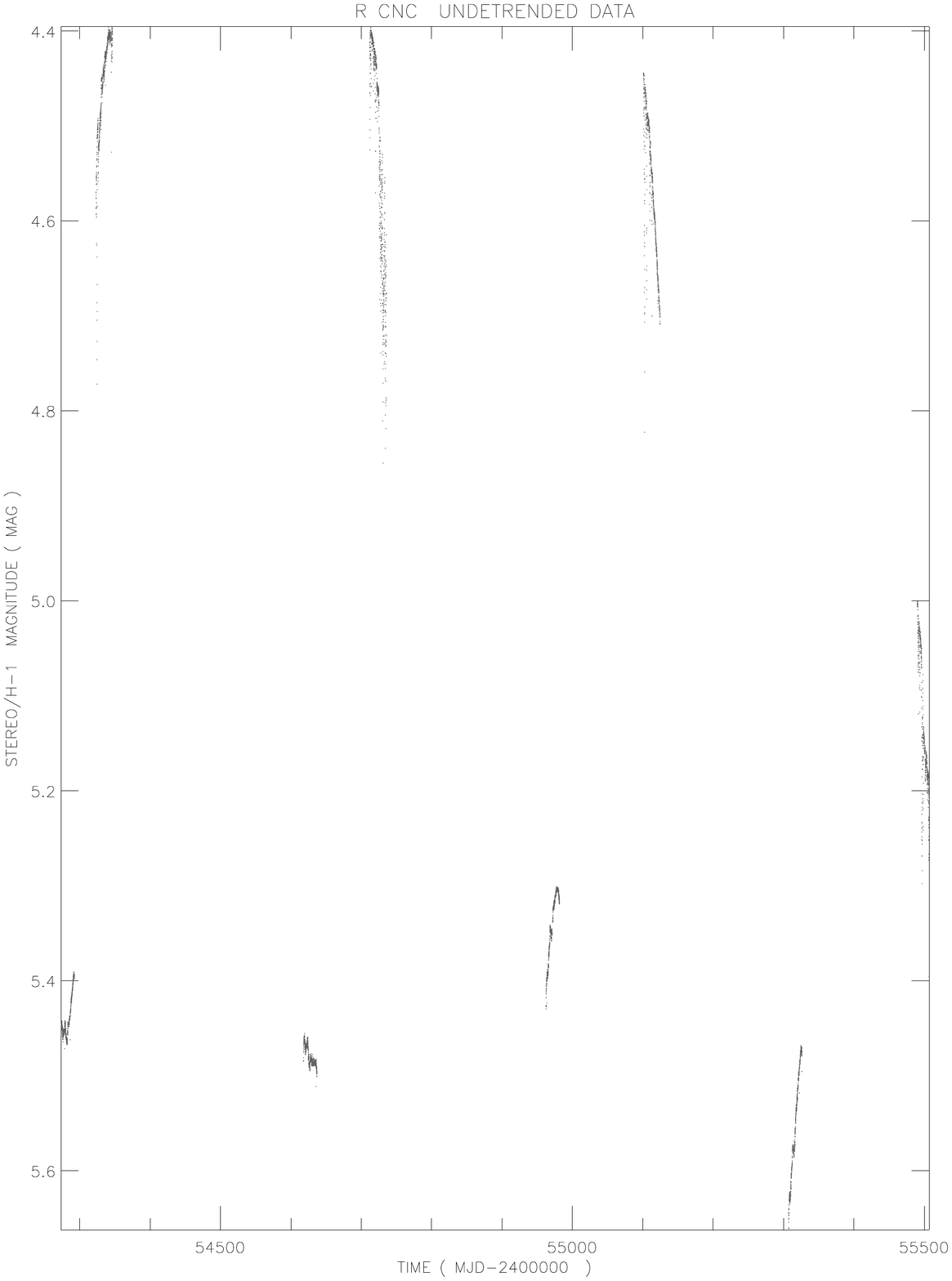}}\hfill
\resizebox{5cm}{!}{\includegraphics{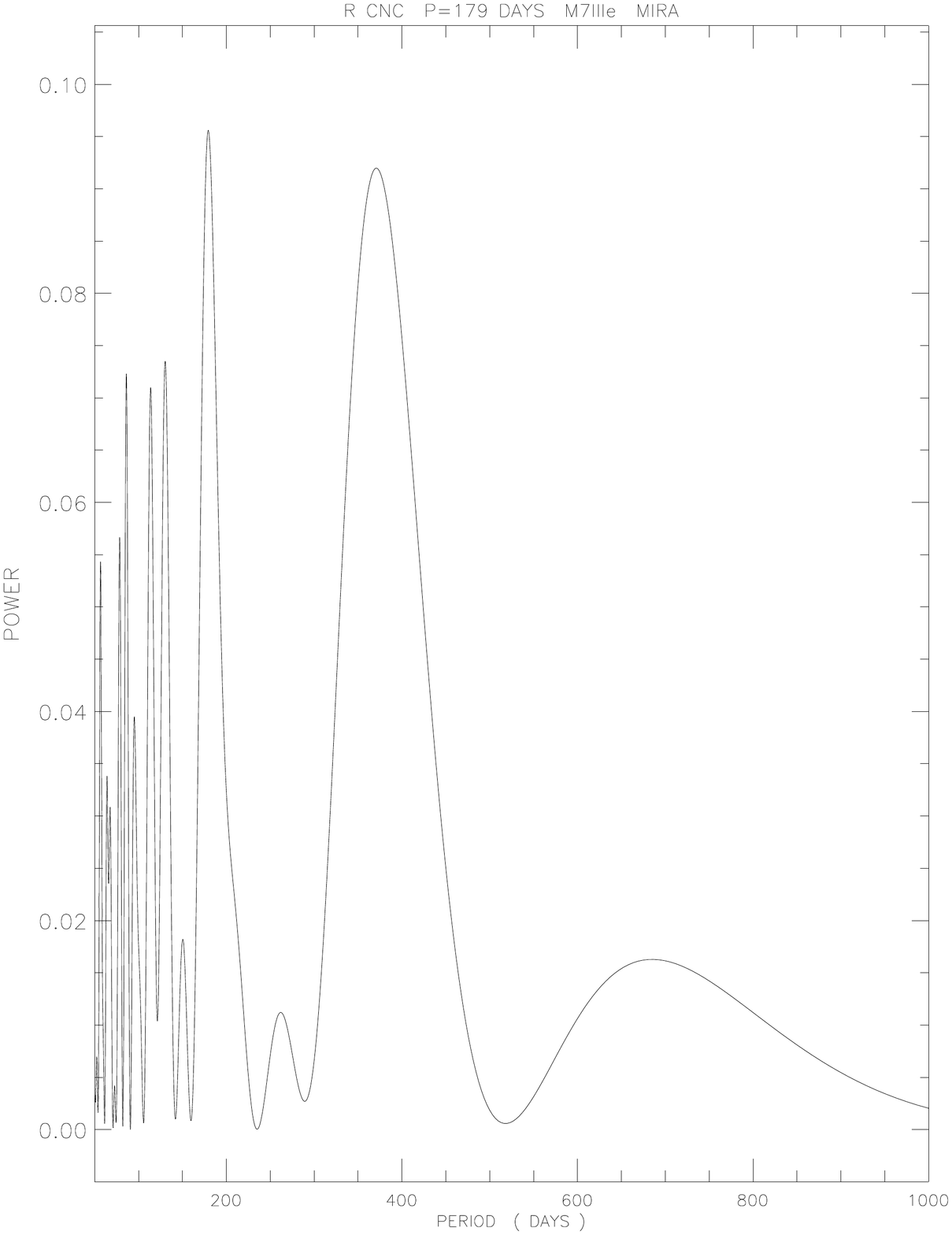}}\hfill
\resizebox{5cm}{!}{\includegraphics{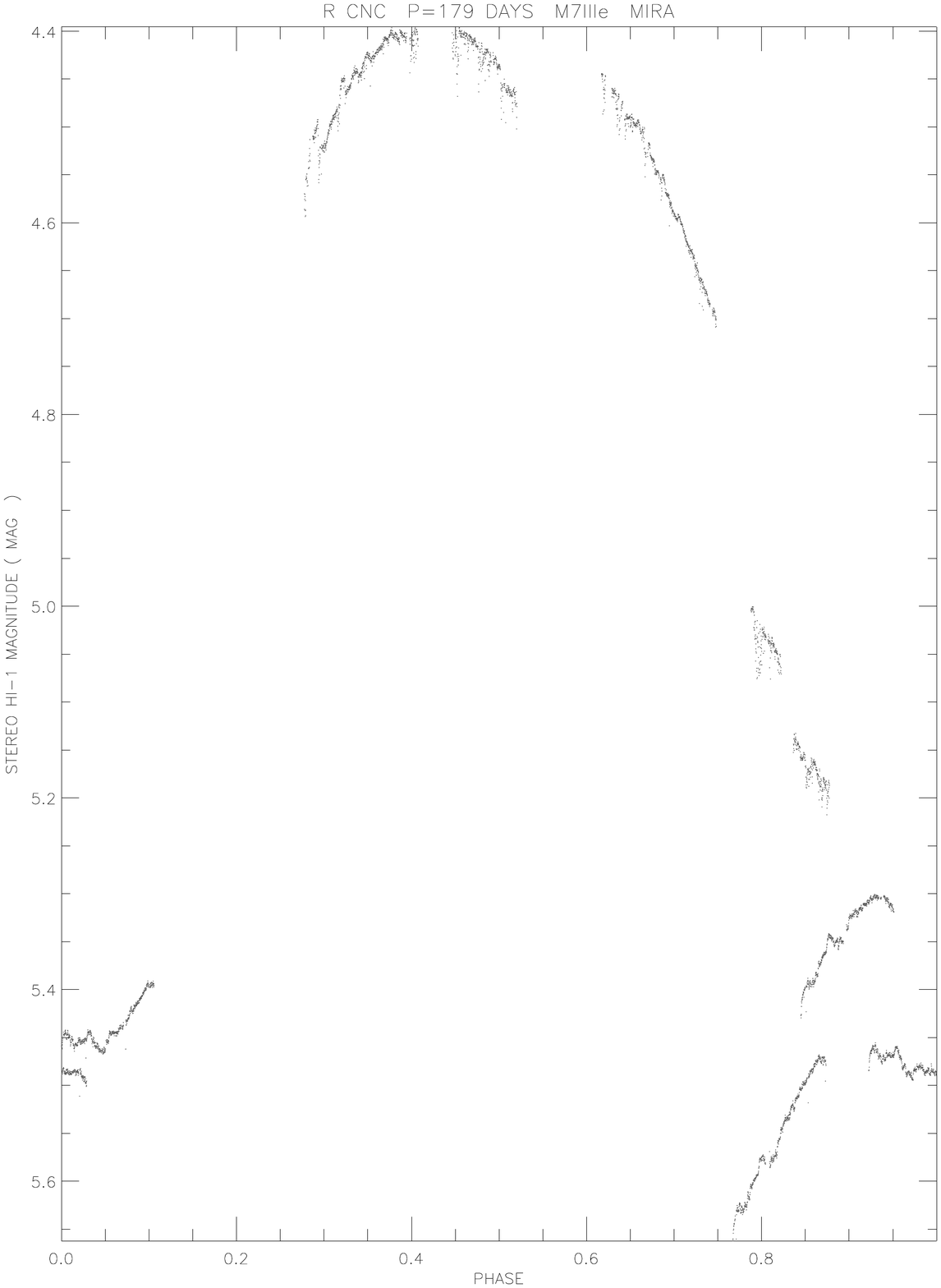}}\hfill
\caption{The known Mira variable R Cnc, as observed by \textit{STEREO}/HI-1.  The raw lightcurve is shown on the left, the DFT of this lightcurve after culling of outliers is then shown in the middle plot and the phase-folded lightcurve on the strongest period (179~days) is shown on the right. Note that the known period of this well-observed star is 361.6~days \protect\citep{samus2012}.}
\label{fig3}
\end{figure*}

\noindent
In order to extract the sample, the first step was to select those stars showing the largest difference in the weighted mean magnitude observed by the two satellites.  We anticipated that the majority of these would be due to systematic effects, mostly relating to flat-fielding near the edges of the CCDs but also due to planetary incursions from Venus and Mercury which frequently pass through the field of view.  The photometry is also conducted very slightly differently between the two imagers, using aperture photometry with different apertures \citep{bewsher2010}.  As a result, there are differences between the weighted mean magnitudes observed by the two satellites, with these systematics being greatest in the Galactic centre and anti-centre.  In normal circumstances, however, the magnitude of these differences amounts to less than 50~mmag and experience has found empirically that systematics are not significant unless the magnitude of the difference between the two satellites is larger than 0.1~mag.  In contrast, the smallest such difference of any object included in the sample is that of V901 Sco, a known semi-regular variable, which in \textit{STEREO}/HI-1 shows a difference of slightly over 0.3~mag between the two satellites and is unlikely to be due solely to systematic effects.  The presence of these systematics is the reason why Miras especially were searched for, as their variability is so large with respect to this background noise that mis-identifications are relatively unlikely.  Approximately 10,000 lightcurves were visually examined showing large differences in the weighted mean magnitudes between the two satellites, from which a sample of about 130 was recovered that appeared to be more likely due to genuine variability than any known systematic effects.  The possibility of this process introducing a selection effect is discussed after the results have been presented, as it is one possible cause for a lack of any objects in the sample showing a periodicity between 200 and 300~days.  For every one of these stars, some basic information was extracted from the \textsc{Simbad} database and the \textsc{NOMAD1} catalogue \citep{zacharias2004}, so as to identify the source of the variability observed.  In many cases, especially in the Galactic centre, it was not possible to be reasonably sure of the origin of the observed variability, mostly because there were too many candidates but often because there were no candidates showing the colours expected of a red giant.  A few were deselected after further examination showed that systematics were more likely responsible for the variability, or because a period could not be found during the final stage of the analysis.  A sample of 85 was eventually analysed in detail using \textsc{Peranso} \footnote{http://www.peranso.com} and a period determination made using a Discrete Fourier Transform \citep{deeming1975} (hereafter DFT), Phase Dispersion Minimisation \citep{stellingwerf1978} (hereafter PDM) and the Renson string length minimisation method \citep{renson1978} (hereafter SLM).  Although \textsc{Peranso} has numerous algorithms implemented, these three have a fundamentally different basis to each other and are expected to have different, hopefully complementary, strengths and weaknesses.  It was important for the methods chosen to be able to deal with a very small number of data points, as each epoch of about 20~days would effectively be treated like a single point.  The ability to straightforwardly and manually remove known, obvious, artefacts, such as planetary incursions due to Venus and Mercury passing through the field of view, made \textsc{Peranso} preferable to other programs, at the expense of being slightly more time-consuming, which was why using numerous additional algorithms was not done.  The median value found by the three algorithms was used, so that if any two algorithms agreed this would produce a more reliable period.  This produced a proportional difference between the periods found and the previously known periods of about 4\verb+%+ (Section \ref{subsec:trends}).  Periods were searched for in the range of 50 to 1000 days and the strongest signal of each of the three different algorithms was recorded, along with the error.  Where possible, times of maxima were also recorded along with the corresponding magnitude observed.  These periods and ephemerides are one of the main results of the research presented herein, along with the photometry presented in the form of lightcurves phase-folded on these periods.  The location of the 85 stars in the sample on the sky is shown in Figure \ref{fig2}.  As an example of a bright well-known Mira, we show the undetrended data and the phase-folded lightcurve for R Cnc along with the periodogram from a DFT analysis of this data in Figure \ref{fig3}.

\section{\uppercase{Results}}
\label{sec:results}

\noindent
The analysis produced periods, with errors, for 85 stars and the time of at least one maximum for 24 of these (Table \ref{table1}) but V932 Sco was excluded from the table as it is a known Orion\footnote{Orion variables are very young stars that have not yet joined the main sequence and show large amplitude irregular variability} variable and the table is likely to be used as a resource for bright LPV stars.  This is the first period determination for 19 of these 85 stars, 7 of which were previously unclassified and 6 previously unknown to be variable.  Table \ref{table2} shows when the 6 previously unknown candidate variables will next be observed by \textit{STEREO}/HI-1.  The phase-folded lightcurves of all 85 stars on the periods found are given in the Appendix (Figures \ref{figA1}, \ref{figA2}, \ref{figA3}, \ref{figA4} and \ref{figA5}).  Unfolded lightcurves for some stars of interest are also shown (see Section \ref{sec:individual}).  V932 Sco has no period given in \citet{samus2012} and the period found here of $423 \pm 87$~days is the first determination of a period, however no maxima were observed.  As it is important to remind the reader that there is a chance that the new variables (marked with \verb+"--"+ in column 11 of Table \ref{table1}) and the unclassified variables (marked as V* in column 11 of Table \ref{table1}) are not necessarily Miras or semi-regular variables, V932 Sco has otherwise been retained in the sample.

Table \ref{table1} below shows the available information for each of the 84 stars (V932 Sco, a known Orion variable, was excluded so as to provide a list of known and probable LPV stars) and lists the name of the known or candidate variable, the name of the star observed by \textit{STEREO}/HI-1 where this is different from the star believed to be the origin of the variability, the right ascension and declination (in degrees) of the star observed by \textit{STEREO}/HI-1 ,the period found from the analysis of the \textit{STEREO}/HI-1 lightcurve (in days), the error given for this period by \textsc{Peranso} (in days), which algorithm produced this period (1 for PDM, 2 for DFT or 3 for SLM), the period given by \textsc{Simbad} (NA is shown if none is available) - preferentially the GCVS \citep{samus2012}, whether the GCVS notes that the period has been observed to vary, the spectral type given by \textsc{Simbad}, the variability type given in the GCVS (including the New Suspected Variable supplement, these are marked as such), where known, followed by the Modified Julian Date (MJD) of all maxima observed in the \textit{STEREO}/HI-1 photometry, a visual estimate of the error in this time (in days) and the \textit{STEREO}/HI-1 magnitude of each maxima.  If a spectral type determination has not been made or the star has not previously been recorded as variable, then \verb+"--"+ is used to denote the absence of this data.

\onecolumn
\scriptsize
\begin{landscape}
\begin{longtable}{|c|c|c|c|c|c|c|c|c|c|c|c|c|c|}
\caption{84 stars showing long period variability in \textit{STEREO}/HI-1 data.  V932 Sco was excluded from the table as this is a known Orion variable and the new and unclassified variables are more likely to be Miras or semi-regular variables, making this table a resource of known and probable LPV stars.} 
\protect\label{table1} \\
\hline
star name & star observed & RA & DEC & \textit{STEREO}/HI-1 & \textit{STEREO}/HI-1 & algorithm & known & period known & spectral type & GCVS & maximum & $\pm$ error in & \textit{STEREO}/HI-1 \\
 & by &  &  & period & period error & 1=PDM & period & to vary &  & variability &  & maximum & magnitude \\
 & \textit{STEREO}/HI-1 & (deg) & (deg) & (days) & (days) & 2=DFT,3=SLM & (days) &  &  & type & (MJD) & (days) & at maximum \\
\hline
 ZCet &&016.6880&-001.4814&186&17&2&184.81&0& M5e & Mira &54604.00&2.5&7.18\\
  &&&&&&&&&  &  &54978.00&4.0&7.60\\
 SPsc &&019.3939&008.9313&413&149&3&404.62&0& M7e & Mira &54606.35&3.5&6.90\\
  &&&&&&&&&  &  &55386.21&4.0&7.44\\
 ZAri &&047.0491&026.9877&341&35&2&337&0& M5 & Mira &54557.77&4.0&8.51\\
 UAri &&047.7627&014.8001&397&48&2&371.1&0& M5.5e & Mira &&&\\
 IKTau &&058.3702&011.4060&449&216&3&470&0& M6me & Mira &&&\\
 V1100Tau & NOMAD1 1102-0050892 &065.3896&020.2805&331&67&3& NA &0& M6me & Mira &&&\\
 V718Tau &&067.8414&017.6529&396&72&1&405&0& Ce & Mira &&&\\
 MYAur &&072.3385&030.9265&323&47&1&331.6&0& -- & Mira &&&\\
 VTau &&073.0096&017.5380&170&8&2&168.7&1& Kp & Mira &&&\\
 EITau &&086.7355&017.9086&360&56&1&364&0& Sv & SR &55280.00&4.0&8.73\\
 ZTau &&088.1039&015.7958&438&81&1&466.2&1& S7.51e & Mira &&&\\
 UOri &&088.9549&020.1752&372&65&1&368.3&1& M8III & Mira &&&\\
 BVAur &&092.7280&030.2310&365&44&2&388&0& M8 & Mira &&&\\
 HVGem & TYC1879-585-1 &097.2217&024.0277&369&68&1&386&0& -- & Mira &&&\\
 RTGem &&101.6440&018.6149&346&21&2&350.4&0& C & Mira &55479.82&4.5&8.03\\
 IUGem & NOMAD1 1100-0135465 &104.3910&020.0554&385&42&2& NA &0& M8e & SR &&&\\
 RGem &&106.8390&022.7035&391&59&2&369.91&0& S2.9e-S8.9e(Tc) & Mira &&&\\
 AMGem & NOMAD1 1183-0161263 &106.7950&028.3008&387&47&2&356.3&0& M10 & Mira &&&\\
 VXGem &&108.2040&014.6010&365&50&2&379.4&0& C7.2e-C9.1e(Nep) & Mira &&&\\
 T CMi &&113.5020&011.7353&325&57&1&328.3&1& M5 & Mira &&&\\
 VWCnc & NOMAD1 1001-016333 &123.4310&010.1629&382&47&2&366&0& M7 & Mira &&&\\
 RCnc &&124.1410&011.7262&179&11&2&361.6&0& M7IIIe & Mira &54978.54&2.0&5.30\\
 WCnc &&137.4690&025.2483&399&50&2&393.22&0& M7e & Mira &&&\\
 SLeo &&167.7120&005.4597&197&25&1&190.16&1& M3me & Mira &55172.05&3.5&9.40\\
  &&&&&&&&&  &  &55358.35&4.0&9.15\\
 TVir &&183.6530&-006.0358&330&113&3&339.47&0& M6e & Mira &&&\\
 SSVir &&186.3100&000.7697&377&55&2&364.14&1& C & SR &55028.79&3.0&5.27\\
  &&&&&&&&&  &  &55370.93&2.5&5.06\\
 SVir &&203.2500&-007.1947&356&56&2&375.1&1& M7IIIe & Mira &&&\\
 KSLib & NOMAD1 0790-0270584 &218.2080&-010.9123&362&46&2&380&0& Me & Mira &&&\\
 EPLib &&219.9990&-022.5740&191&19&1&185.78&0& -- & Mira &&&\\
 SXLib & HD129380 &220.6990&-020.2238&324&129&3&332.9&0& M6e & Mira &&&\\
 EGLib &&223.8400&-022.0055&402&378&3&365&0& M5 & Mira &&&\\
 SLib &&230.3500&-020.3884&197&24&1&192.9&1& M2 & Mira &&&\\
 SVLib &&233.3410&-027.1855&405&101&1&402.66&0& M8+ & Mira &&&\\
 TZLib &&234.2140&-010.0802&193&20&1&183.6&0& -- & Mira &54462.38&3.0&8.86\\
  &&&&&&&&&  &  &54844.66&4.0&8.72\\
 XSco &&242.1330&-021.5305&198&34&1&199.86&0& M2 & Mira &55245.09&3.0&9.02\\
 WWSco & NOMAD1 0587-0415805 &246.8190&-031.2615&435&86&1&431&0& M9 & Mira &&&\\
 YSco & NOMAD1 0706-0340442 &247.3880&-019.3839&355&48&2&351.88&1& M8 & Mira &54474.10&2.5&8.82\\
  &  &&&&&&&&  &  &55255.82&4.0&8.45\\
 WXSco &&248.2020&-026.3850&385&85&1&187.9&0& M6 & Mira &&&\\
 TOph &&248.4310&-016.1317&377&45&2&366.82&0& M6.5e & Mira &&&\\
 XZSco &&249.2640&-027.3132&365&50&2&300&0& -- & Mira &&&\\
 YYSco & NOMAD1 0614-0400462 &249.5170&-028.5985&379&106&1&327.43&0& M7e & Mira &&&\\
 CISco & NOMAD1 0602-0420519 &250.5110&-029.7392&389&109&1& NA &0& -- & V* &&&\\
 IRAS 16482-2039 & NOMAD1 0692-0384552 &252.8140&-020.7219&376&50&2& NA &0& -- & -- &&&\\
 IRAS 16469-3211 & NOMAD1 0577-0577145 &252.5500&-032.2835&360&48&2& NA &0& -- & V*(NSV) &&&\\
 CROph & NOMAD1 0611-0442400 &253.8260&-028.8998&327&74&1&345.7&0& M & Mira &&&\\
 V1163Oph &&254.6800&-016.8690&321&49&1&324&0& -- & Mira &&&\\
 EGOph & NOMAD1 0639-0431340 &254.7280&-026.0260&410&79&2& NA &0& -- & V* &&&\\
 V901Sco &&255.6910&-032.7255&409&112&3& NA &0& Ne & SR &&&\\
 GPOph &&256.2530&-027.2164&328&83&1& NA &0& M6 & SR &54871.96&3.5&8.73\\
 ROph &&256.9410&-016.0927&302&127&3&306.5&1& M4e & Mira &55256.71&2.0&5.70\\
 AEOph & NOMAD1 0699-0414346 &259.4700&-020.0211&359&49&2&176&0& -- & Mira &55444.80&5.0&8.21\\
 IRAS 17289-1917 & NOMAD1 0706-0429226 &262.9790&-019.3255&171&10&1& NA &0& -- & -- &&&\\
 BGOph & NOMAD1 0732-0506383 &265.4140&-016.7940&386&56&2&342.5&0& M9 & Mira &&&\\
 IRC -30357 & NOMAD1 0635-0722217 &271.8500&-026.4031&353&50&2& NA &0& M8 & -- &&&\\
 BRSgr &&275.0510&-032.2159&309&61&1&302.8&0& M4e & Mira &&&\\
 V1869Sgr &&275.9300&-033.2451&318&35&2&332&1& Me & Mira &&&\\
 AKSgr &&277.005&-016.7509&385&37&2&413.15&0& M5e-M9 & Mira &&&\\
 HRSgr &&277.0920&-028.2411&380&46&2& NA &0& -- & V* &&&\\
 V3876Sgr &&278.3030&-020.0973&344&87&1&352&1& M8 & Mira &54890.57&3.0&8.02\\
 IRC -20507 & NOMAD1 0714-0715901 &280.0740&-018.5605&431&86&1& NA &0& M7 & -- &55464.71&5.0&8.55\\
 V3867Sgr & NOMAD1 0595-0911163 &280.7240&-030.5022&414&183&3&422&0& -- & Mira &&&\\
 V3878Sgr & NOMAD1 0618-1015599 &280.7590&-028.1300&357&45&2&345&0& -- & Mira &&&\\
 V3952Sgr &&281.4640&-017.2999&492&153&2& NA &0& M9 & Mira &55287.07&5.0&7.61\\
 V2055Sgr &&283.0480&-032.8287&321&34&1&320&0& -- & Mira &&&\\
 OPSgr &&283.3480&-026.3368&397&47&2&303&0& Me & Mira &&&\\
 V5545Sgr & NOMAD1 0676-0984820 &283.4830&-022.3865&368&46&2&377&0& Me & SR &54506.99&2.5&8.50\\
  & &&&&&&&&  &  &54895.71&3.0&8.35\\
  & &&&&&&&&  &  &55283.38&5.0&8.52\\
 NSV11552 &&284.0560&-017.7138&179&16&2& NA &0& -- & V* &&&\\
 FQSgr &&286.9030&-017.0215&432&87&1&434&0& M8 & Mira &55285.02&5.0&6.89\\
 RXSgr &&288.6370&-018.8120&326&87&1&335.23&1& M5e & Mira &54780.85&3.0&7.23\\
 TYSgr &&289.4280&-023.9402&324&126&3&325.41&0& M3e & Mira &&&\\
 ANSgr &&291.7610&-018.5139&325&53&3&337.56&1& M5e-M8 & Mira &&&\\
 IRAS 19263-1922 &&292.3270&-019.2722&386&41&2& NA &0& -- & V*(NSV) &&&\\
 2MASS J19291709-2034504 & NOMAD1 0693-0875859 &292.3310&-020.6250&425&87&1& NA &0& M7 & -- &&&\\
 V360Sgr &&293.9280&-012.7919&367&46&2&165&0& M7 & SR &54522.44&3.0&9.39\\
  &&&&&&&&&  &  &54905.91&3.0&9.28\\
  &&&&&&&&&  &  &55295.30&5.0&9.71\\
 NOMAD1 0784-0674630 &&298.3360&-011.5768&381&51&2& NA &0& -- & -- &&&\\
 RRSgr &&298.9850&-029.1900&335&131&2&336.33&1& M5e & Mira &&&\\
 IRAS 20060-2425 &&302.2470&-024.2704&376&376&3& NA &0& -- & V*(NSV) &&&\\
 RCap &&302.8260&-014.2676&175&10&2&345.13&0& Cev & Mira &54524.99&3.0&7.53\\
 SWCap &&306.5840&-009.5489&363&51&2&344.6&0& M8 & Mira &&&\\
 XXAqr &&310.5790&-008.2585&345&37&2&323.4&0& M4 & Mira &&&\\
 ZCap &&317.6560&-016.1737&181&17&2&181.48&0& M Iab:e & Mira &&&\\
 RZAqr &&320.7650&-007.1082&429&117&1&391&0& M9 & Mira &54549.49&3.5&7.09\\
 HYAqr &&322.7770&-007.5723&309&65&1&311&0& M8 & Mira &54819.52&3.5&8.86\\
 XAqr &&334.6640&-020.9011&306&58&1&311.4&0& S6.3e M4e-M6.5e & Mira &&&\\
\hline
\end{longtable}

\begin{longtable}{|cccc|ccccc|}
\caption{The dates shown (in MJD-2400000) in this table are when the 6 new candidate variables will next be observed by \textit{STEREO}/HI-1.  The times take into account the mask that is routinely applied to exclude data likely to be affected by solar activity.  As the Sun is going to be at maximum during these observations, it is not expected that observations will be reliable within the region of the CCDs excluded by the mask.} 
\protect\label{table2} \\
\hline
star name & star observed & RA & DEC & \multicolumn{5}{|c|}{start and end dates of upcoming \textit{STEREO}/HI-1 observations (MJD-2400000)}\\ \cline{5-9}
& by \textit{STEREO}/HI-1 & (deg) & (deg) & \textit{STEREO}/HI-1A & \textit{STEREO}/HI-1B & \textit{STEREO}/HI-1A & \textit{STEREO}/HI-1B & \textit{STEREO}/HI-1A\\
\hline
IRAS 16482-2039 & NOMAD1 0692-0384552 &252.8140&-020.7219& 56119.30339 - 56131.47135 & 56418.86914 - 56432.20117 & 56463.46094 - 56475.62891 & 56807.82813 - 56821.16016 & 56807.61849 - 56819.78646 \\
IRAS 17289-1917 & NOMAD1 0706-0429226 &262.9790&-019.3255& 56129.12500 - 56142.40234 & 56428.35156 - 56441.51953 & 56473.65234 - 56486.92969 & 56817.29688 - 56830.46484 & 56818.17969 - 56831.45703  \\
IRC -30357 & NOMAD1 0635-0722217 &271.8500&-026.4031& 56137.03385 - 56150.17057 & 56436.99805 - 56450.05664 & 56481.57031 - 56494.70703 & 56826.03906 - 56839.09766 & 56826.10677 - 56839.24349  \\
IRC -20507 & NOMAD1 0714-0715901 &280.0740&-018.5605& 56144.72005 - 56158.02474 & 56444.58789 - 56457.72852 & 56489.23828 - 56502.54297 & 56833.54688 - 56846.68750 & 56833.75651 - 56847.06120  \\
2MASS J19291709-2034504 & NOMAD1 0693-0875859 &292.3310&-020.6250& 56155.52474 - 56168.82943 & 56456.15234 - 56469.20703 & 56500.04297 - 56513.34766 & 56845.20703 - 56858.26172 & 56844.56120 - 56857.86589 \\
NOMAD1 0784-0674630 &&298.3360&-011.5768& 56162.64714 - 56182.14714 & 56456.62109 - 56476.50781 & 56507.15625 - 56526.65625 & 56845.66406 - 56865.55078 & 56851.66536 - 56871.16536  \\\hline
\end{longtable}
\end{landscape}

\normalsize
\twocolumn

\begin{figure}
\resizebox{2.6cm}{!}{\includegraphics{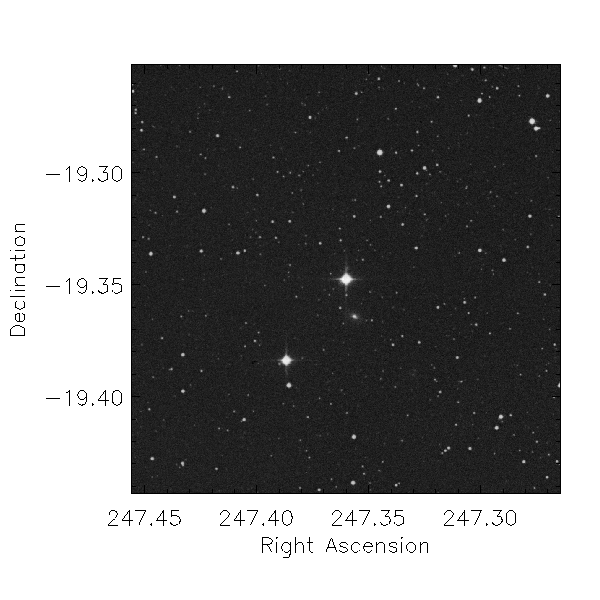}}
\hfill
\resizebox{2.6cm}{!}{\includegraphics{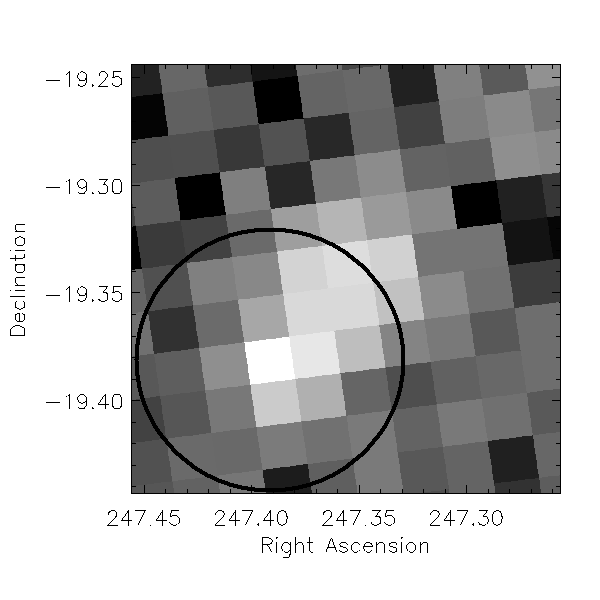}}
\hfill
\resizebox{2.6cm}{!}{\includegraphics{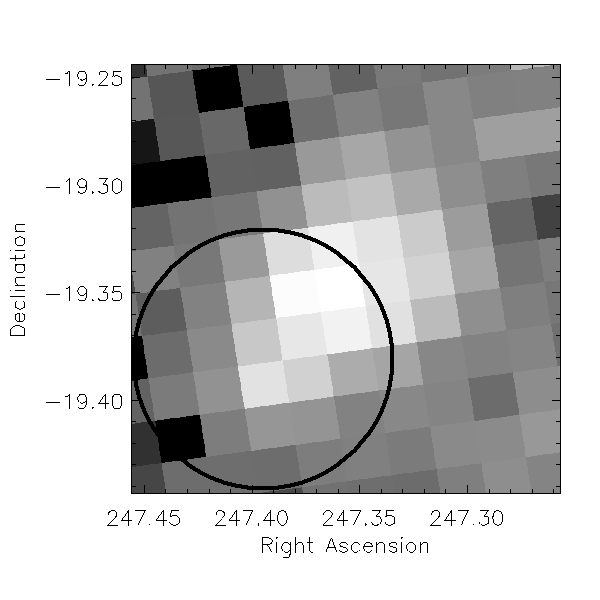}}
\caption{ESO \textit{R} band image centred on the Mira variable Y Sco (left), \textit{STEREO}/HI-1A image showing the same field of view near minimum (middle) and \textit{STEREO}/HI-1B image of the same field of view at maximum (right).  The nearby bright star is NOMAD1 0706-0340442 and is the only star in this field of view in the \textit{STEREO} database.  The variability of Y Sco was detected and its period determined by virtue of the blending effect with this star, about 2.4 pixels away.  The overlaid dark circles illustrate the photometric aperture of the Heliospheric Imagers, which is slightly different for each imager, at 3.2 pixels for \textit{STEREO}/HI-1A and 3.1 pixels for \textit{STEREO}/HI-1B \protect\citep{bewsher2010}.  The galaxy with a bright core just below Y Sco has not been recorded.}
\label{fig4}
\end{figure}

The sample of 85 stars for which a period was determined were analysed for trends in their periods.  Some of this analysis was done using \textsc{R}\footnote{http://www.R-project.org}\citep{rproject} and some using \textsc{topcat}\footnote{http://www.starlink.ac.uk/topcat/}.  The main reason for doing this was that the errors returned for the periods by \textsc{Peranso} were often very large and it was necessary to verify whether they were valid.  Secondly, there were no stars in the sample for which the best period found was between 200 and 300~days, which was not expected.  All these 85 stars were also checked for known periods in \textsc{Simbad}, the General Catalogue of Variable Stars (GCVS) \citep{samus2012} period being used preferentially where multiple determinations were available.   The actual \textit{STEREO}/HI-1 images of several stars were examined showing maximum and minimum brightness in order to ascertain that even from a distance of 2 or 3 pixels, some of the stars in the sample could still have been observed indirectly (e.g. Y Sco, shown in Figure \ref{fig4}).

\subsection{Trends in the data}
\label{subsec:trends}

\begin{figure}
\resizebox{8cm}{!}{\includegraphics{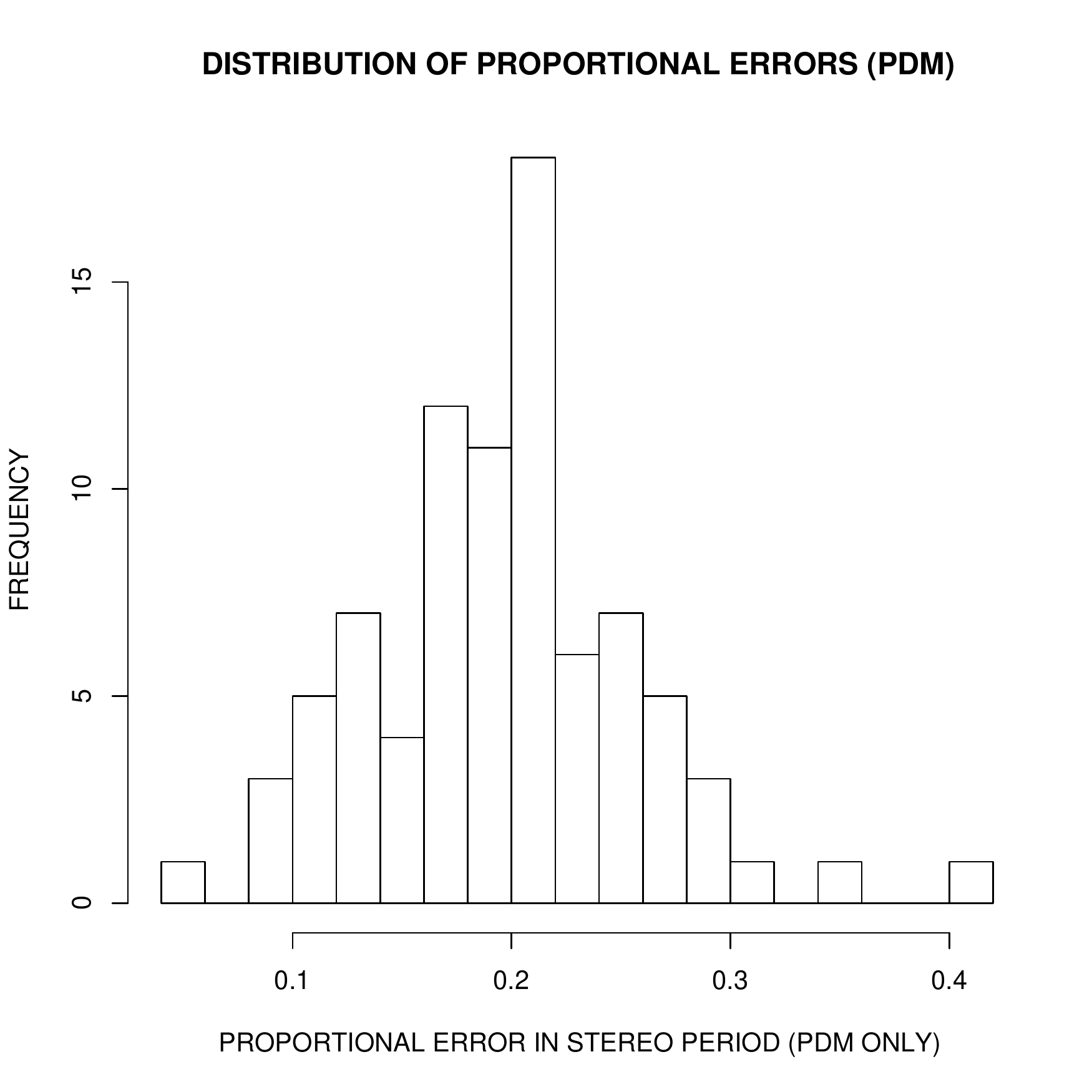}}
\caption{Histogram showing the distribution of the proportional errors in the periods observed by \textit{STEREO}/HI-1 using Phase Dispersion Minimisation \protect\citep{stellingwerf1978}.}
\label{fig5}
\end{figure}

\begin{figure}
\resizebox{8cm}{!}{\includegraphics{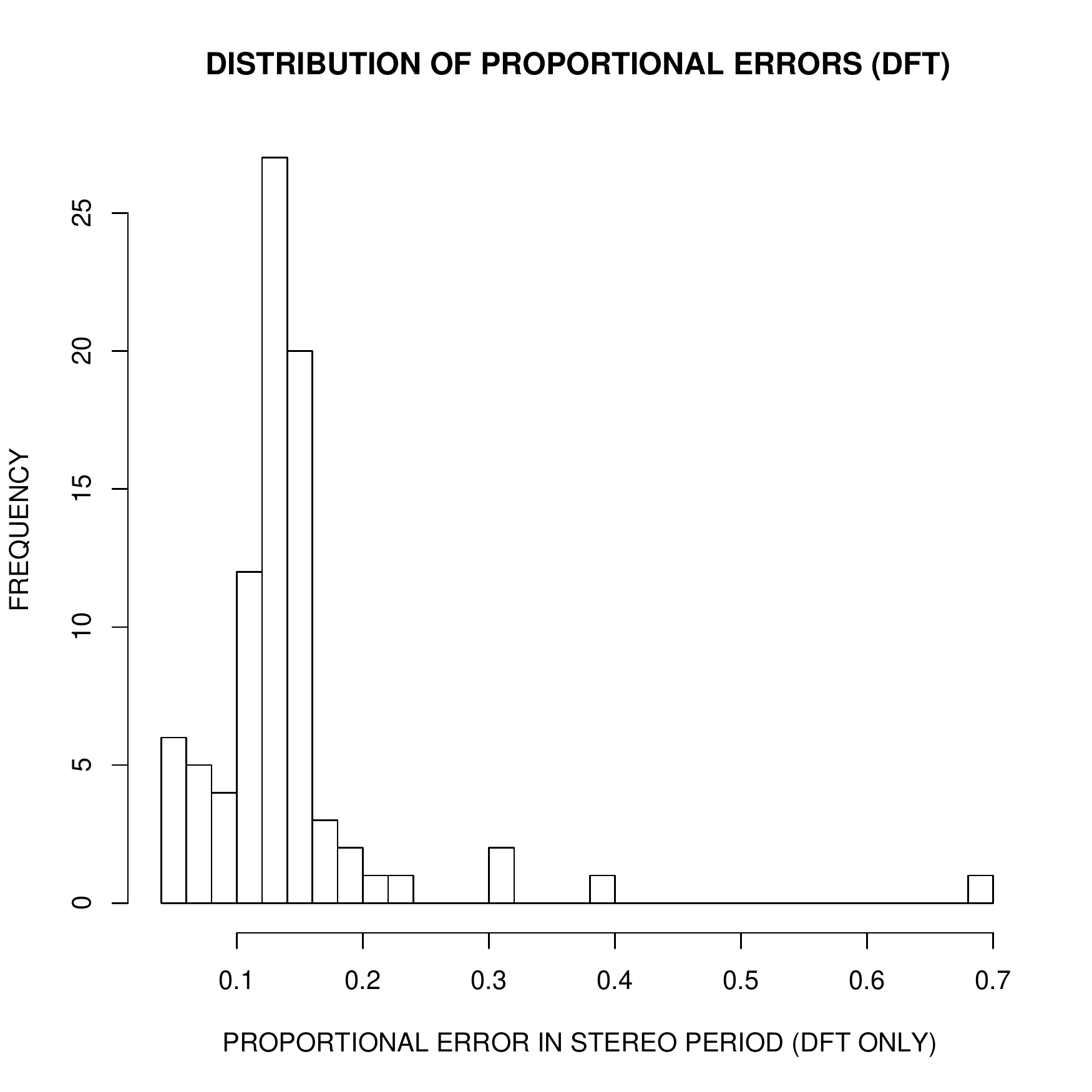}}
\caption{Histogram showing the distribution of the proportional errors in the periods observed by \textit{STEREO}/HI-1 using a Discrete Fourier Transform \protect\citep{deeming1975}.}
\label{fig6}
\end{figure}

\begin{figure}
\resizebox{8cm}{!}{\includegraphics{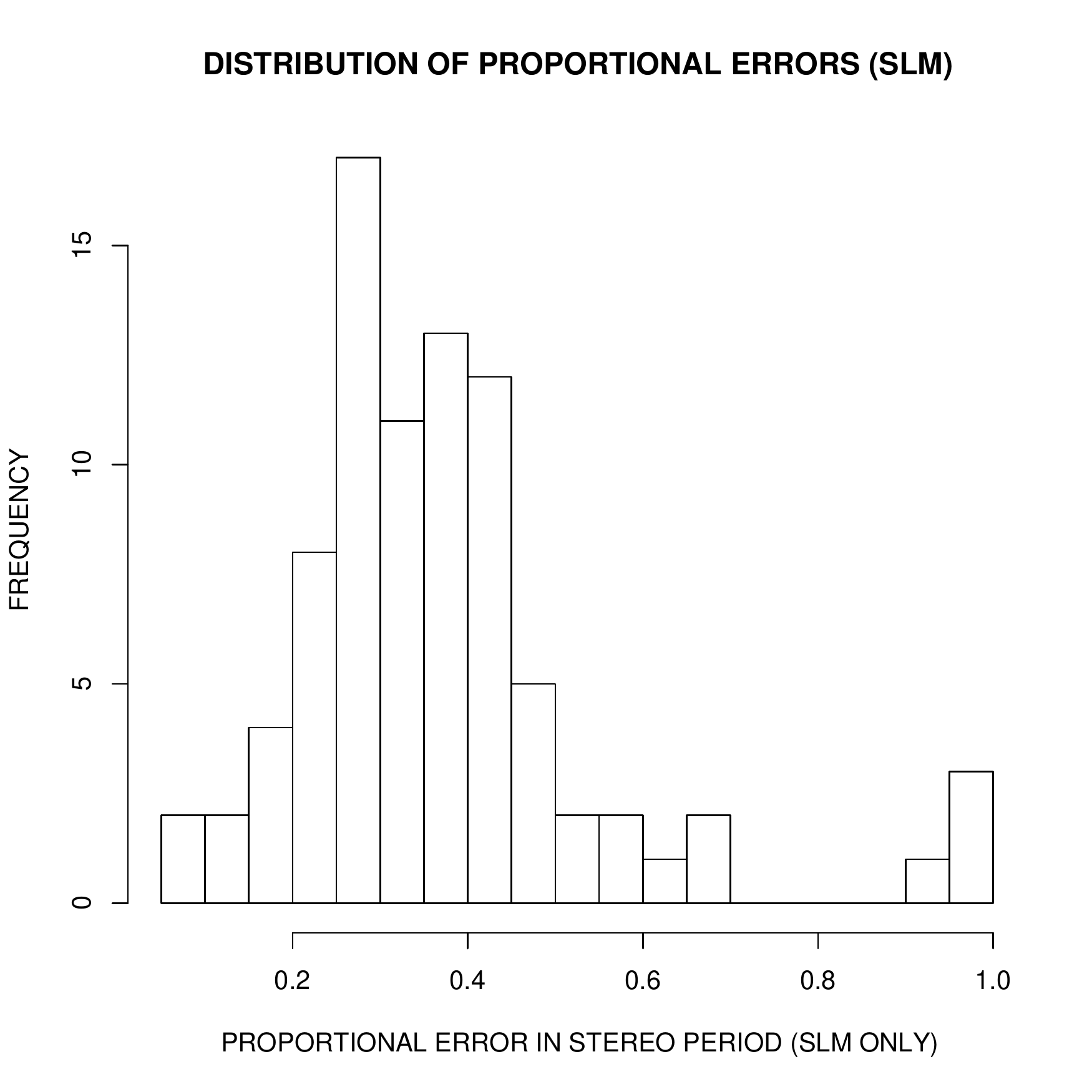}}
\caption{Histogram showing the distribution of the proportional errors in the periods observed by \textit{STEREO}/HI-1 using String Length Minimisation \protect\citep{renson1978}.}
\label{fig7}
\end{figure}

\begin{figure}
\resizebox{8cm}{!}{\includegraphics{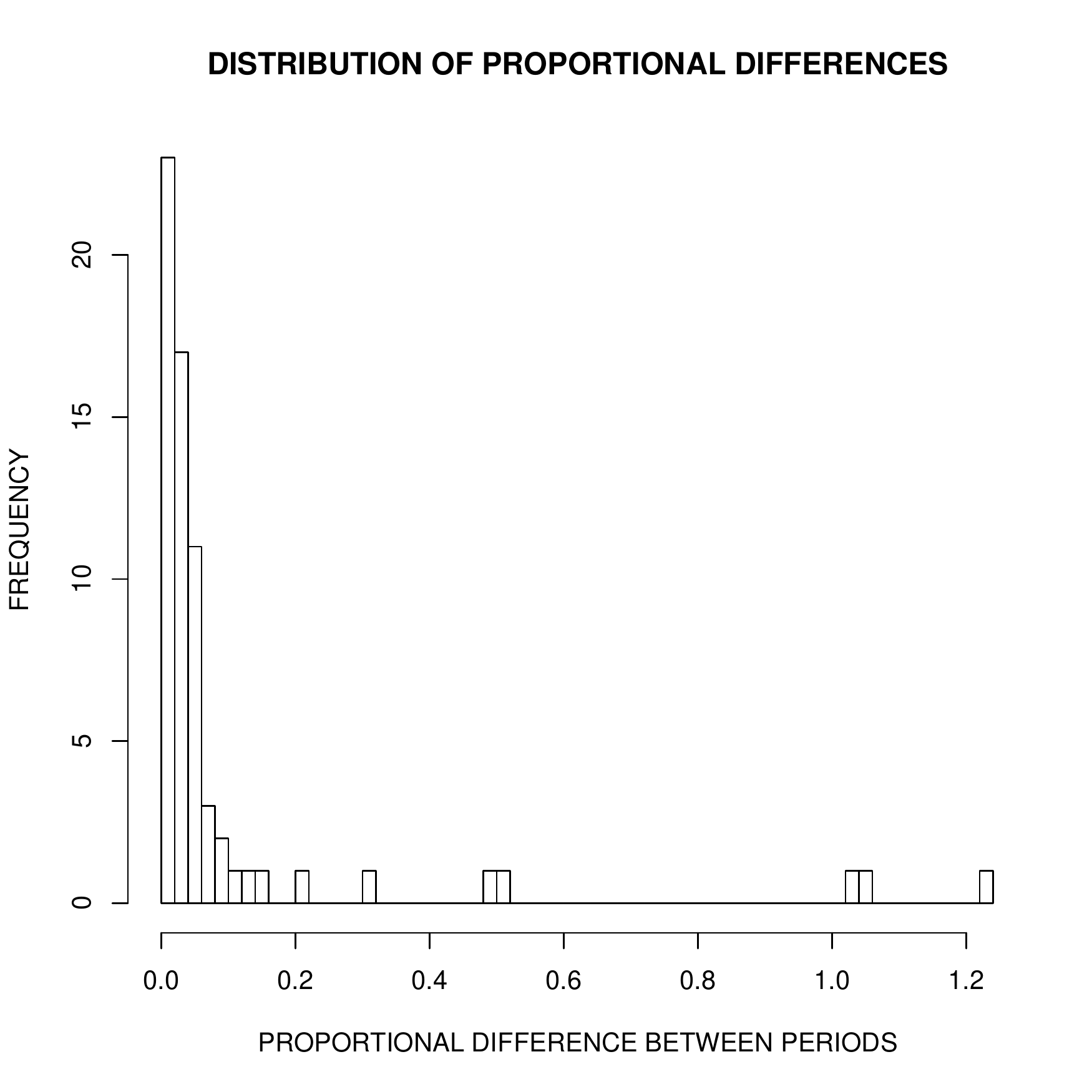}}
\caption{Histogram showing the distribution in the proportional differences ($(P_{STEREO}-P_{GCVS})/P_{GCVS}$) between the \textit{STEREO}/HI-1 periods and those found in the literature \protect\citep[preferentially][]{samus2012}.}
\label{fig8}
\end{figure}

\begin{figure}
\resizebox{8cm}{!}{\includegraphics{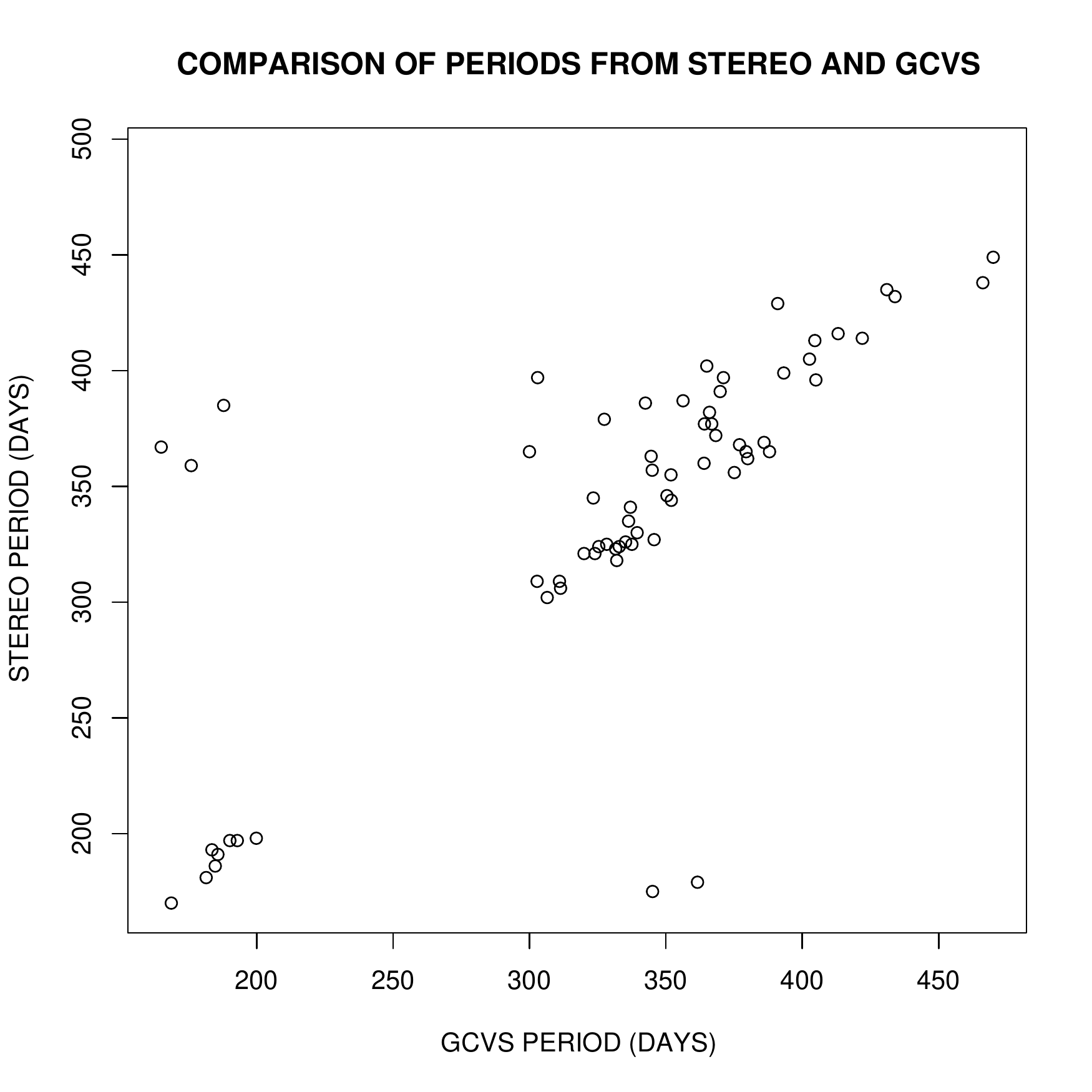}}
\caption{Plot comparing the periods observed by \textit{STEREO}/HI-1 to those found in the literature \protect\citep[preferentially][]{samus2012}.}
\label{fig9}
\end{figure}

\noindent
The concern regarding the accuracy of the errors in the periods returned by \textsc{Peranso} is demonstrated by Figures \ref{fig5}, \ref{fig6} and \ref{fig7}.  The large size of the errors implies that the algorithms were, individually, struggling to resolve a signal, however by comparing the median period of the three with the known periods from the GCVS \citep{samus2012} shows that a more appropriate 1$\sigma$ error bar for the entire sample is 4\verb+%+ (Figure \ref{fig8}).  A direct plot of the median period found here against the GCVS periods shows a few cases where harmonics may have been found by \textit{STEREO}/HI-1 instead of the correct period but confirms the overall good match (Figure \ref{fig9}).  The table giving all the median periods and other information of relevance for each star in the sample is found in the appendix in Table \ref{table1}.  Lightcurves phase-folded on these periods are also found in the appendix.

\begin{figure}
\resizebox{4cm}{!}{\includegraphics{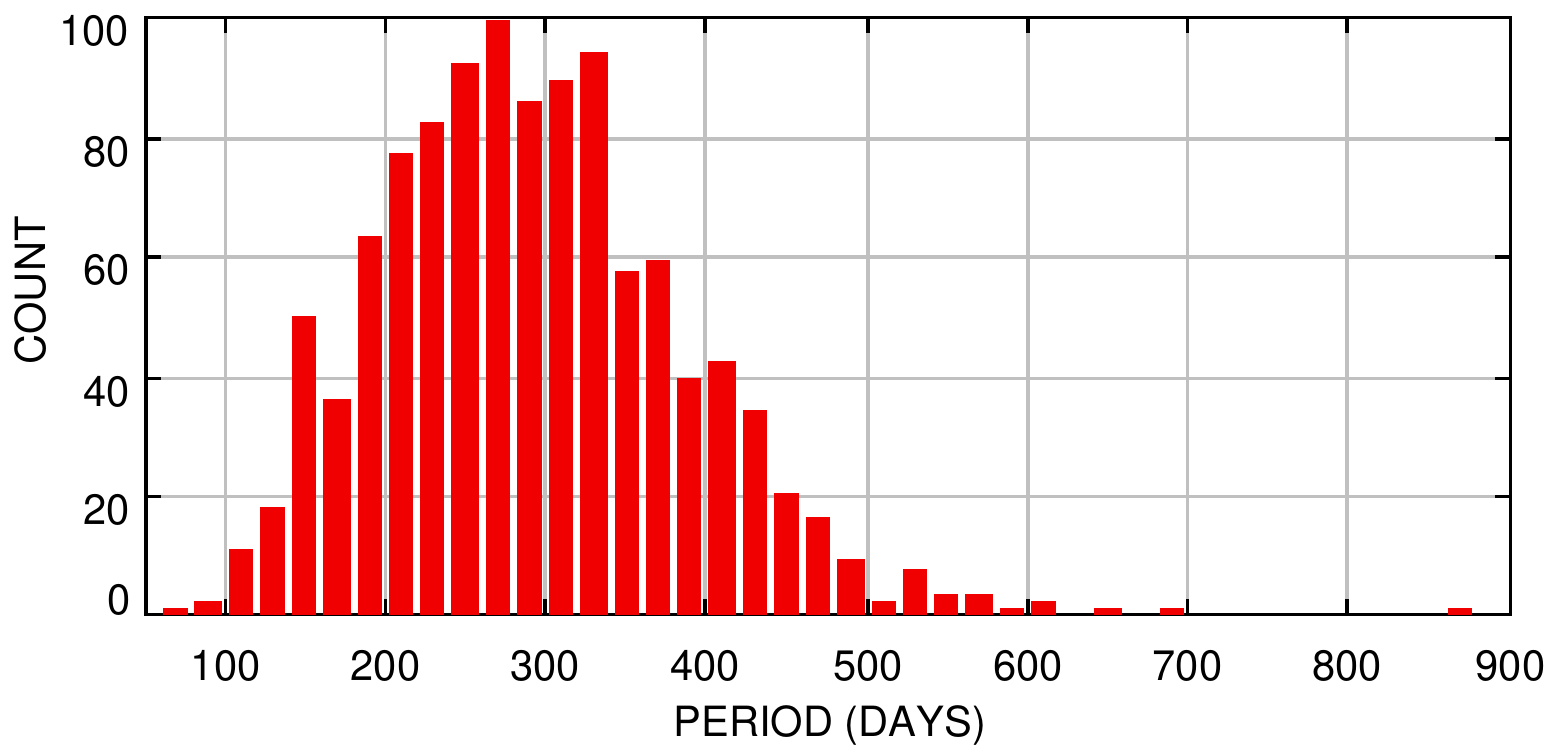}}
\hfill
\resizebox{4cm}{!}{\includegraphics{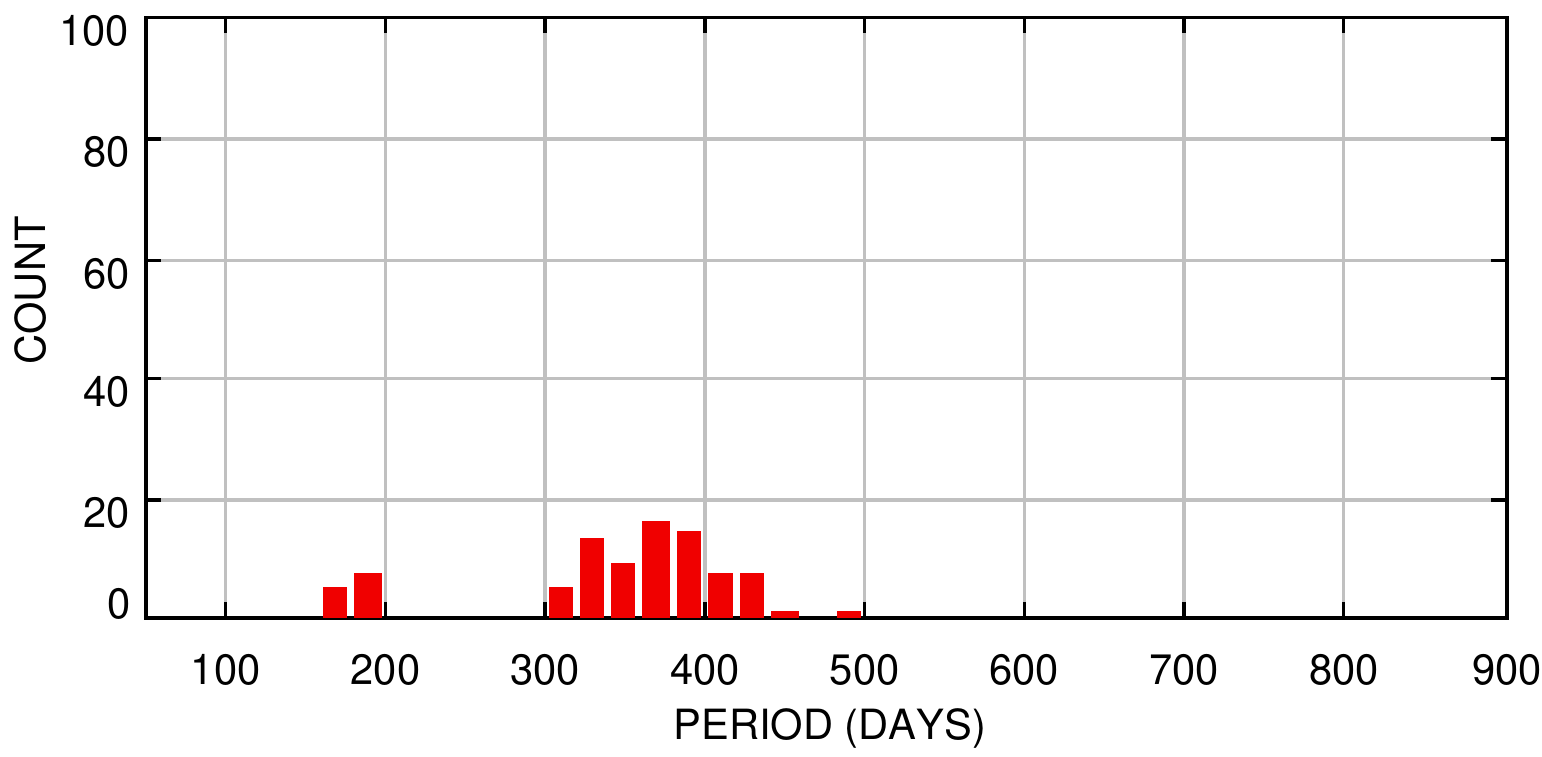}}
\caption{Histograms comparing the periods of known Mira variables from \protect\citet{kharchenko2000}, shown on the left, with the distribution of periods found for all variables in this work (right).  The lack of periods between 200 and 300 days in this work is believed to be due to a selection effect, with periods in this range being more difficult to distinguish from systematic effects.}
\label{fig10}
\end{figure}

There are no periods found between 200 and 300 days in the sample, as shown on the right in Figure \ref{fig10}.  This is in contrast to what is expected, as shown by the distribution of the known periods of Miras from \citet{kharchenko2000} displayed on the left in Figure \ref{fig10}.  This is believed to be the result of the way the sample was selected, with a visual examination of large numbers of lightcurves: periods in this range would have a less distinctive pattern in the lightcurve and be almost impossible to distinguish from systematic effects.

\begin{figure}
\resizebox{4cm}{!}{\includegraphics{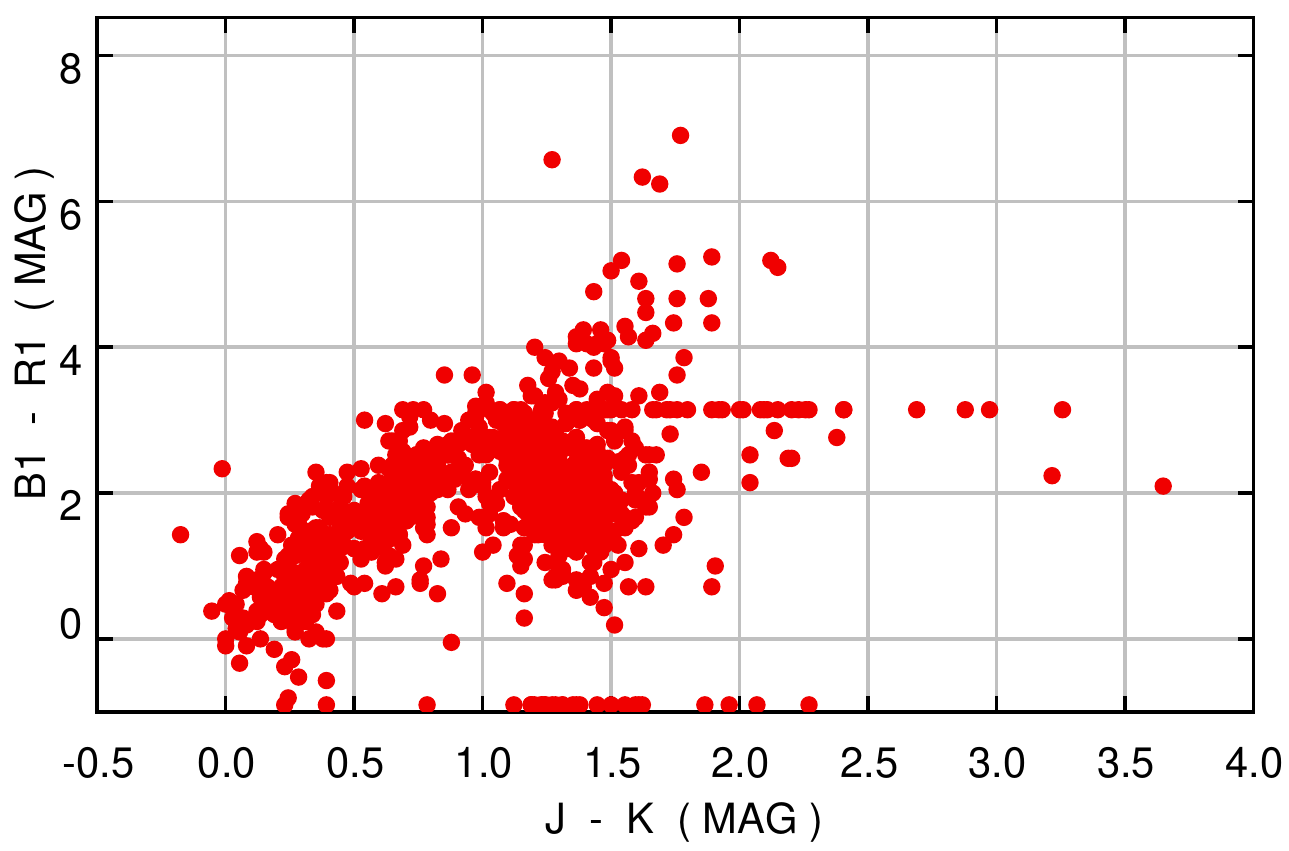}}
\hfill
\resizebox{4cm}{!}{\includegraphics{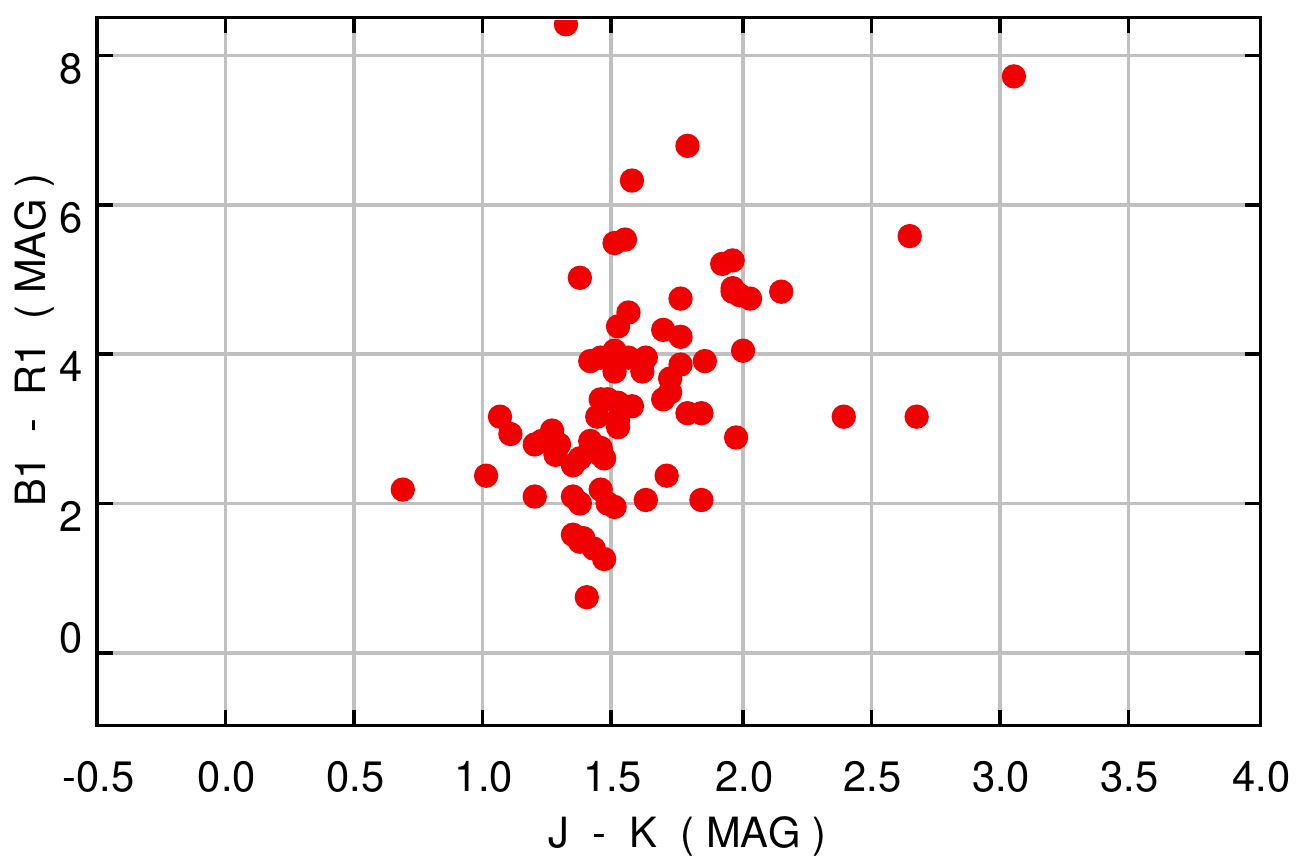}}
\caption{Plot of the $B1-R1$ colours \protect\citep{monet2003} against $J-K$ colours \protect\citep{skrutskie2006} for all the Miras in \protect\citet{kharchenko2000}, left, and all the stars in this sample observed by \textit{STEREO}/HI-1, right.  V2055 Sgr, a known Mira, is the star with the largest value of $B1-R1$ in our sample, whilst V718 Tau, also a known Mira, has the second-largest value of $B1-R1$ and the largest value of $J-K$.}
\label{fig11}
\end{figure}

\begin{figure}
\resizebox{4cm}{!}{\includegraphics{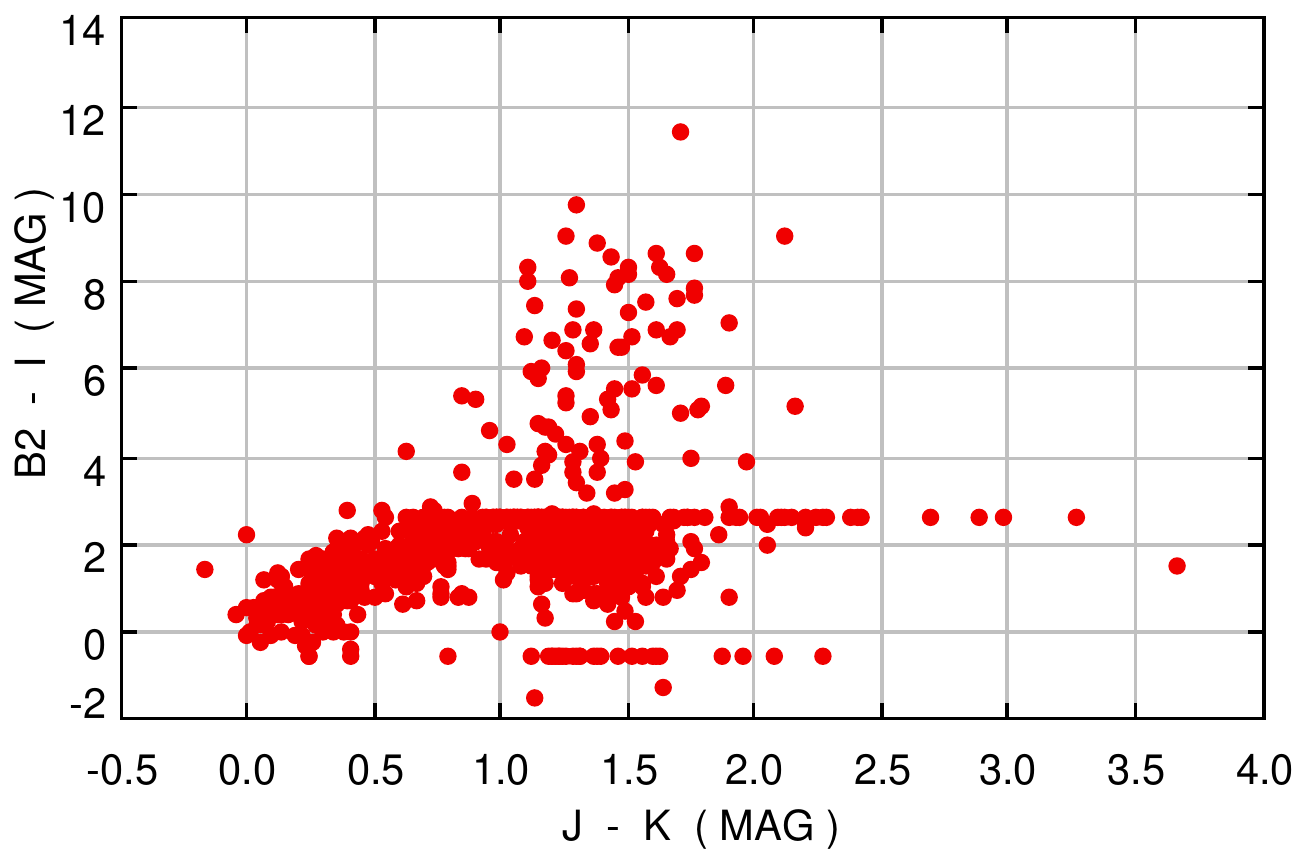}}
\hfill
\resizebox{4cm}{!}{\includegraphics{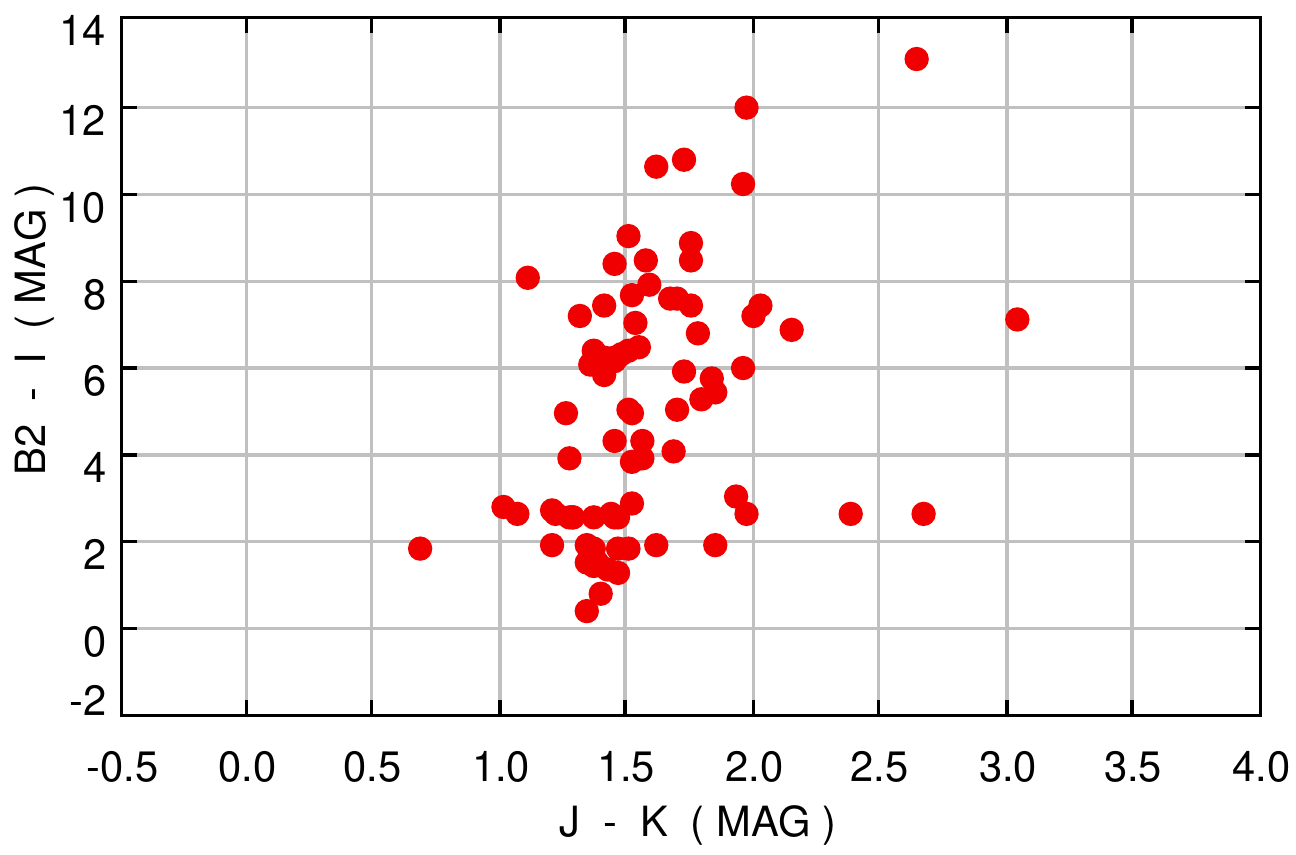}}
\caption{Plot of the $B2-I$ colours \protect\citep{monet2003} against $J-K$ colours \protect\citep{skrutskie2006} for all the Miras in \protect\citet{kharchenko2000}, left, and all the stars in this sample observed by \textit{STEREO}/HI-1, right.  The stars with $(B2-I)>10$ from the \textit{STEREO}/HI-1 sample are IK Tau, IRC -20507, 2MASS J19291709-2034504, OP Sgr and FQ Sgr, in decreasing order.}
\label{fig12}
\end{figure}

The distribution of colours of known Miras from \citet{kharchenko2000} shows two main populations (Figure \ref{fig11}, left) when comparing the $B1-R1$ colours from \citet{monet2003} against the $J-K$ colours from \citet{skrutskie2006}.  The distribution of colours of stars in this sample observed by \textit{STEREO}/HI-1 is very different, however (Figure \ref{fig11}, right).  This may in part reflect that it includes some semi-regular variables but might also indicate that some of the stars in the sample have unusual features, such as circumstellar dust shells, or that they  could have been misclassified.  Checking the $B2-I$ colours against $J-K$, from the same sources, shows a population of objects (Figure \ref{fig12}, left) which is a better match for the sample observed by \textit{STEREO}/HI-1 (Figure \ref{fig12}, right).  A slight bias for new and unclassified objects to have large $B2-I$ is expected as a result of checking for Mira-like colours (i.e. very red objects) when attempting to ascertain the source of variability, however it was not expected that the known Miras found would share this feature.  The throughput of the \textit{STEREO}/HI-1 imagers at about 950~nm is more likely to be responsible, as these objects are bright in the \textit{I} band. 

\section{\uppercase{Notes on individual stars}}
\label{sec:individual}

\noindent
This section details \textit{STEREO}/HI-1 observations of particular stars of individual interest.    To clarify the observations of those stars without a prior period determination, unfolded lightcurves are given for all the new candidates (Figure \ref{fig13}), all the previously unclassified variables (Figure \ref{fig14}) and the remaining stars for which a period has not previously been determined (Figure \ref{fig15}).  A few general comments about the observations are also worth noting:
\begin{itemize}
\item The times of maxima observed do not always correspond to the periods given in Table \ref{table1}.  This is because the times of maxima, the shape and amplitude of the lightcurve can all potentially change from cycle to cycle and the algorithms do not always give a good fit.
\item The shape of the lightcurve at maximum can sometimes be very flat, such that the star appears constant for many days, even longer than a single epoch of observations (about 20~days).  In these cases, no maximum was recorded as it was not possible to determine a central time of maximum with an estimate of the error.  Therefore, only those maxima sharply defined enough to produce an error of about 5~days or less were recorded.
\item In cases where the \textit{STEREO}/HI-1 period found is a harmonic of the known period, it is difficult to know which is genuine, especially for those with periods near to a year or a fraction thereof.  It would be be necessary to combine all available observations to be sure for these stars but, although this is beyond the scope of the present paper, due to the relatively small number of data points herein we are not questioning the accuracy of periods given in the literature.
\item Classifications for the new long period variable candidates, or previously unclassified variables, cannot be accurately determined as the magnitude of variability in the non-standard \textit{STEREO}/HI-1 bandpass is not straightforward to convert to a more standard filter.  In comparison to the brighter, better-known Miras, however, it is likely that most are semi-regular variables, if genuine.  The reddening for many of the new candidates, some recorded as having circumstellar envelopes, is an additional complication.
\item For 19 stars in the sample, the period presented here constitutes the first determination of a period.  The irregular nature of these variables and the large errors in the periods require that these are taken with caution, individually, although for those stars with a known period there is general good agreement with the periods found here and thus there should also be similarly good agreement, overall, with those for which no period has previously been reported.
\end{itemize}

\subsection{New candidate and unclassified variables}

\begin{figure}
\resizebox{16cm}{!}{\includegraphics{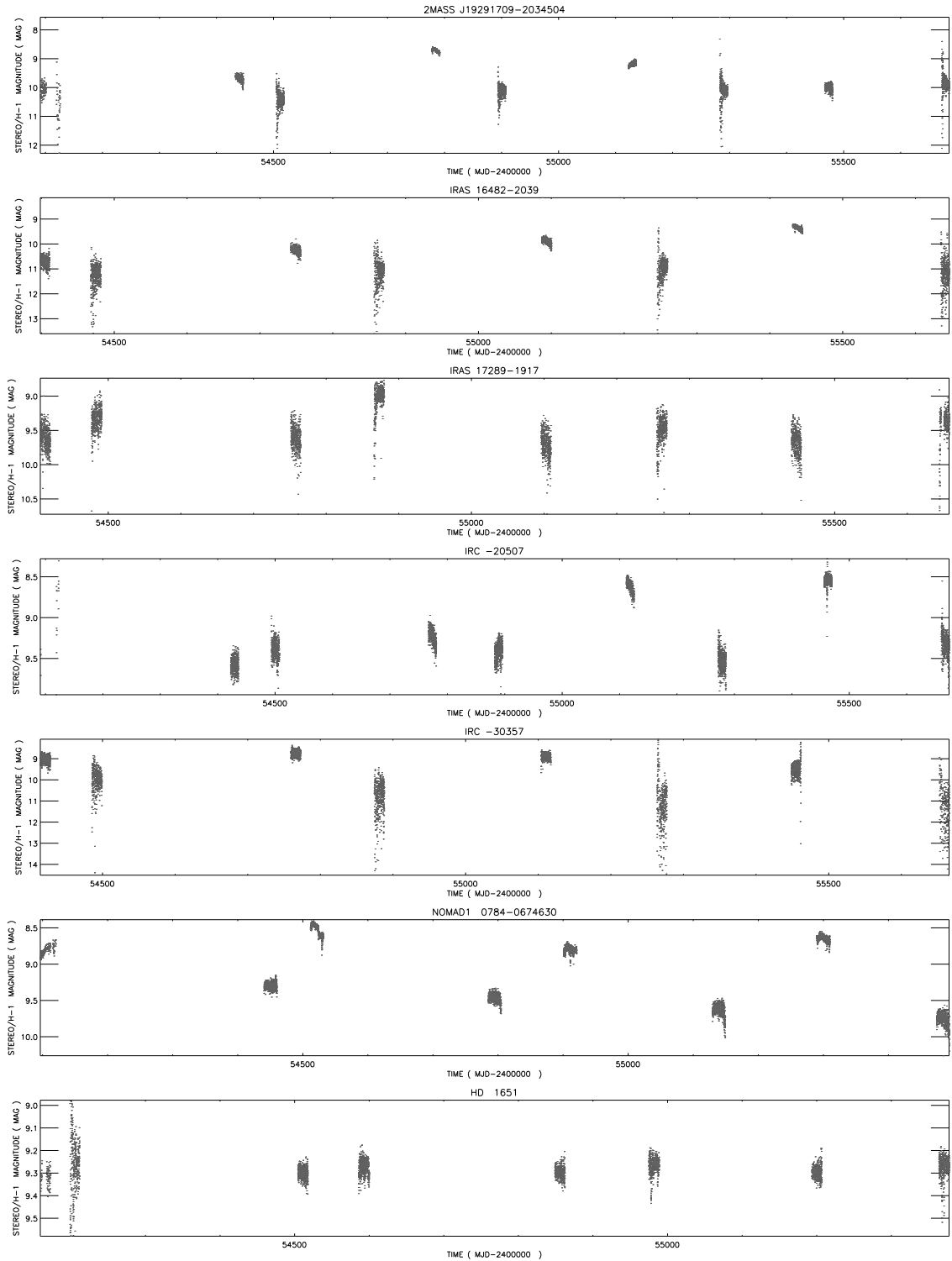}}
\caption{Unfolded lightcurves of the new candidate long period variables.  The data shown here was extracted for their co-ordinates as given in \textsc{Simbad} for those detected indirectly through the pipeline described in Section \ref{sec:analysis} ( all except NOMAD1 0784-0674630 ) and a constant star for comparison ( HD 1651 ).  The clear variability seen here further confirms that such indirect detections are able to recover the large amplitude signals of Miras and semi-regular variables.  Note that the constant star shows a small discrepancy between the \textit{STEREO}/HI-1A and \textit{STEREO}/HI-1B data but this is not on the scale of variability seen in the stars in our sample of long period variables and it also remains constant in each satellite's data, unlike the stars in our sample.}
\label{fig13}
\end{figure}

\noindent
There are 6 candidate long period variables in the sample not listed as variables in the GCVS or its NSV supplement \citep{samus2012} and a further 7 are listed as variables but not classified.  Owing to the difficulties in determining the amplitude of variability, it is not possible to ascertain whether these are Miras or semi-regular variables, however given their apparent magnitudes and the lack of prior measurements or detection of their variability, it is perhaps more likely that most are semi-regular variables.  In particular, the sample was selected partly by the colours shown in \citet{zacharias2004} and are very red objects, thus the likely classifications are limited to Miras and semi-regular variables.  As five of the new candidates were detected indirectly through blending with a nearby star, their lightcurves were also individually extracted using the co-ordinates provided by \textsc{Simbad}, in order to confirm the new sources are accurate.  The lightcurves of all the new candidates are thus presented in Figure \ref{fig13}, along with a constant star for comparison (HD~1651), demonstrating that large amplitude variability has genuinely been observed in these stars. Note that the constant star has a small difference in the observed magnitudes between \textit{STEREO}/HI-1A and \textit{STEREO}/HI-1B but that this is on the order of 50~mmag and that it remains constant for each satellite.

The new candidate long period variables are:
\begin{description}
\item[IRAS 16482-2039]  There are probable systematic effects confusing the behaviour near minima.  Nevertheless, the amplitude of variability clearly exceeds one magnitude.  Only the region near minimum brightness and the decline from maximum are observed, as to be expected for a star with a period close to the orbital period of the \textit{STEREO}-Behind satellite.  This star is recorded as an infra-red source by \textit{IRAS} \citep{neugebauer1984} and has no mention of variability.  It also appears in the \textit{AKARI} FIS All-sky survey point source catalogue \citep{yamamura2010} and has fluxes at 65, 90 and 140 microns, the flux at 140 microns being equivalent to a magnitude of about 7.6, with the other two bands registering a magnitude of about 9.9.  It was observed to show SiO masing in \citet{deguchi2004}, which implies that it is an evolved O-rich star undergoing mass-loss.
\item[IRAS 17289-1917]  This star has either sharp maxima or maxima of varying brightness between epochs, although the minima appear smooth.  The period is very close to one year and any variability could thus easily have been difficult or impossible to observe prior to the \textit{STEREO}/HI-1 observations.  It is recorded as an IR source by \textit{IRAS} and features in \citet{skrutskie2006} but it does not appear in \citet{yamamura2010} and has not been the subject of further investigation.
\item[IRC -30357]  This star shows extremely large amplitude variability but is hampered by the fact that it encroaches on the faint limit observable by \textit{STEREO}/HI-1.  It is possible that only the maxima are seen and that the star is too faint to observe at other times.  The brightness of maximum may be variable between cycles.  The star has been observed by \textit{IRAS} and has been assigned a spectral type of M8, although there is no mention of variability and it does not appear in \citet{yamamura2010}.  The \textit{STEREO}/HI-1 data also shows other variability with a period visually near 1 day, probably a result of blending with a nearby star possessing rotational variability, although none are recorded as such.
\item[IRC -20507]  The phase-folded lightcurve is relatively smooth for this star but it is unclear if there is a change in the brightness of maxima and minima between cycles.  One maximum is observed but it is possible a second was just caught in the previous cycle, in which case the period given here of 431~days may be very wrong, with the most recent cycle having about 352~days between adjacent maxima.  This would then be a sign of irregular variability or a significant period change.  OH masing has been observed in this star \citep{Hekkert1991OHmaser1612} and we therefore expect it to have a circumstellar envelope, making it a particularly interesting target for future study.  The large value of $B2-I$ \citep{monet2003}, second only to IK Tau, is suggestive of the obscuring presence of a circumstellar envelope.  The star is designated spectral type M7 in \textsc{Simbad}.  In \citet{yamamura2010} this star has fluxes at 65 and 90 microns, with equivalent magnitudes of 6.4 and 7.15, respectively.  Variability with a periodicity visually near 1 day is also seen in the data, likely due to blending with a rotational variable, although none are recorded as such nearby.
\item[2MASS J19291709-2034504]  No maxima are observed for this star but there are indications that the brightness of maximum could be variable.  The star is listed in \textsc{Simbad} as spectral type M7 but very few observations have been made and no indication of variability is mentioned.  The extremely large value of $B2-I$ \citep{monet2003} is the third-largest in the sample, after IK Tau and IRC -20507, which suggests that this star may host a circumstellar envelope.  Follow-up observations to check for maser activity would be required to confirm this.
\item[NOMAD1 0784-0674630]  This star shows an erratic unfolded lightcurve and the median period of 381~days indicates the shape and brightness of maxima and minima may be variable.  There were no candidate sources near the observed co-ordinates other than this star that are likely to be Miras, with two other sources very nearby in \citet{zacharias2004}, of which one has colours resembling an A-type star and the other a large proper motion typical of a red dwarf.  Another red star almost 3 arc-minutes away is less likely to be the source, TYC 5733-2876-1 (NOMAD1 0784-0674475), owing to its distance, magnitude and the magnitude of variability observed, although it should not be excluded from consideration.
\end{description}

\begin{figure}
\resizebox{16cm}{!}{\includegraphics{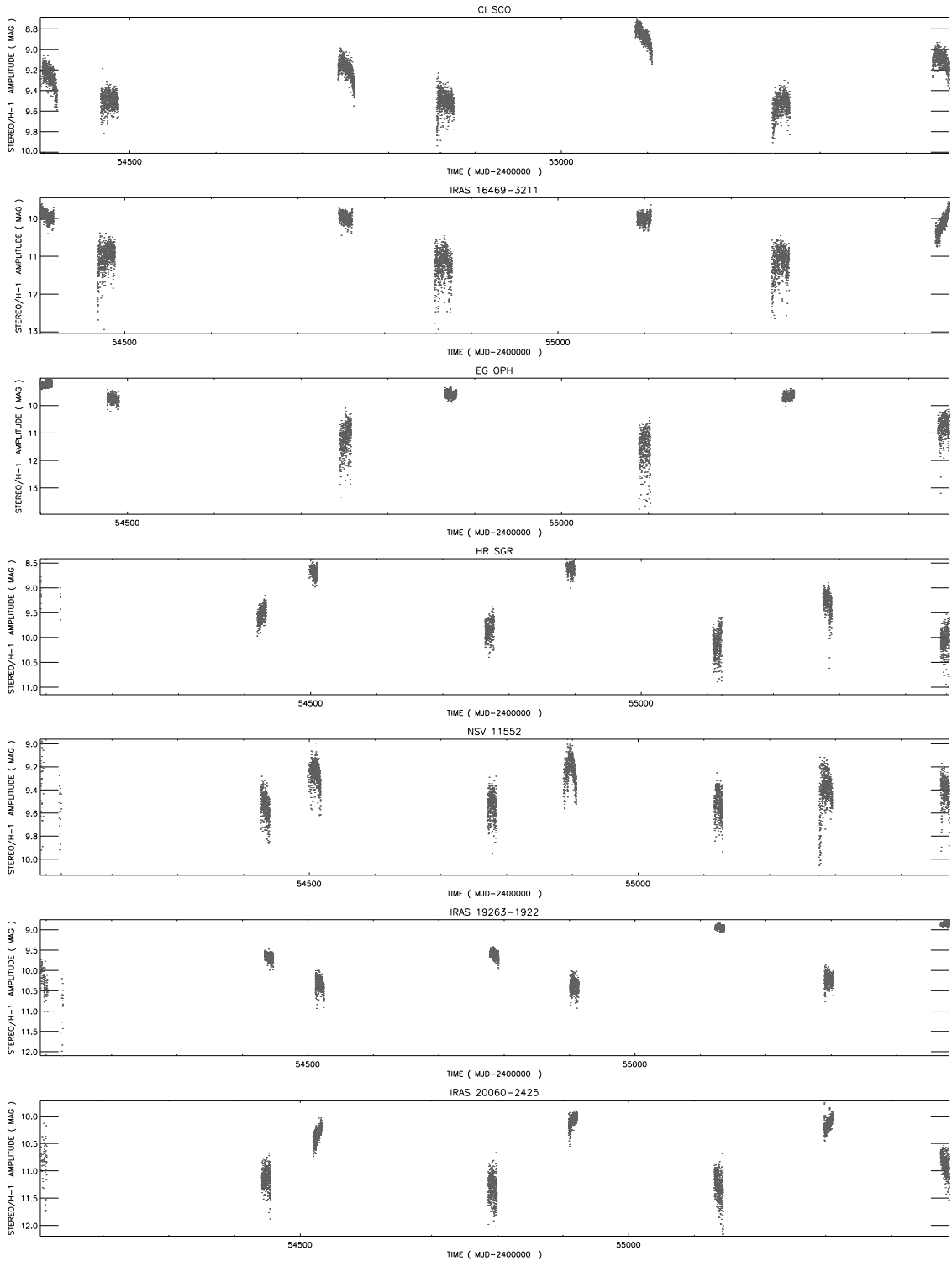}}
\caption{Unfolded lightcurves of the unclassified long period variables.}
\label{fig14}
\end{figure}

\begin{figure}
\resizebox{16cm}{!}{\includegraphics{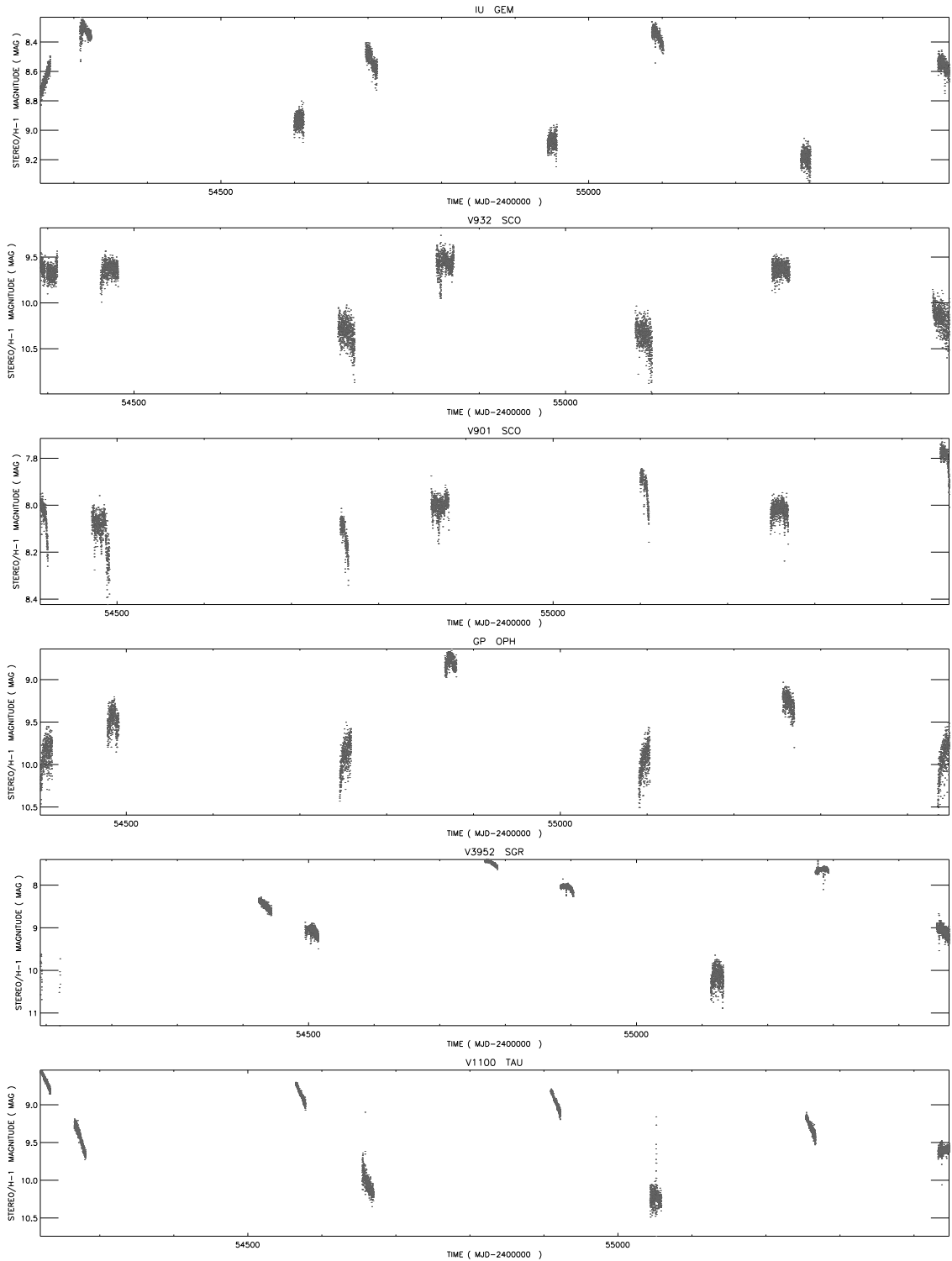}}
\caption{Unfolded lightcurves of the previously known variables for which the period found here is the first to be determined.}
\label{fig15}
\end{figure}

The previously unclassified variables (their unfolded lightcurves are shown in Figure \ref{fig14}) are:
\begin{description}
\item[CI Sco]  The \textit{STEREO}/HI-1A data appears to show a systematic trend that may be potentially confusing the period determination, although all the \textit{STEREO}/HI-1B data appear to be at or very near minimum, consistent with the period being close to the orbital period of the \textit{STEREO}-Behind satellite.  There is some other variability observed in the lightcurve with a period visually near 5~days, although it is not clear in which star this originates.  Although the variability is unclassified and no period is given, an amplitude of 1.5~magnitudes has been recorded \citep{samus2012}, which is about twice that observed by \textit{STEREO}/HI-1.
\item[IRAS 16469-3211]  Poor phase coverage for this star contributes to an uncertainty in both the period and, given the uncertainties in the differences in the magnitudes seen by the two satellites, to the overall amplitude.  One epoch of data clearly shows a sharp increase in brightness and the phase-folded lightcurve places this close to the maximum in phase.  This star is recorded as having an envelope of OH/IR type as a result of observations reported in \citet{Hekkert1991OHmaser1612} in which OH masing was detected.  It was previously observed by \textit{IRAS} but there is no mention of variability.  The actual source observed by \textit{STEREO}/HI-1 is nearby (NOMAD1 0577-0577145) and has very red colours also and there is a chance that this star may be the variable, if it is itself a Mira (however it has no known variability of its own or observations other than photometry).
\item[EG Oph]  For this star the phase-folded lightcurve is smooth except for the first epoch of \textit{STEREO}/HI-1A data, which might indicate a brighter maximum for that cycle.  No period or classification has been determined for this star but the amplitude is reported as 1.4~magnitudes \citep{samus2012}.  It has been observed to be a SiO maser source \citep{deguchi2004}.  There is a risk of contamination from EI Oph, approximately 3 pixels away in \textit{STEREO}/HI-1, which has the same amplitude but is fainter and even more poorly-observed.
\item[HR Sgr]  No period or classification is given in the GCVS or NSV \citep{samus2012}, however the given amplitude of 4.2~magnitudes is suggestive of a Mira.  The amplitude observed by \textit{STEREO}/HI-1 of about 1.5~magnitudes is not unusual in the context of the known Miras in this sample.  There are indications that the signal seen is blended with a variable resembling a WUMa type eclipsing binary, although this would not be expected to produce variability on the scale of a Mira.  The median period found here is close to the orbital period of the \textit{STEREO}-Behind satellite but the smoother parts of the phase-folded lightcurve are from the \textit{STEREO}-Ahead satellite so a systematic is less likely than changing brightness of the maxima between cycles.
\item[NSV 11552]  This star has the lowest amplitude of variability in the sample and also one of the shortest periods.  The amplitude is given in \textsc{Simbad} as 1.7~magnitudes, however, so this is unlikely to be rotational variability.  It is unclear from the \textit{STEREO}/HI-1 data whether the period might instead be twice that given here.  Some other variability due to blending is evident in the lightcurve, although the source is uncertain.  The two brightest stars in \textit{R} nearby have significant proper motions indicative of red dwarf stars rather than Miras.
\item[IRAS 19263-1922]  The \textit{STEREO}/HI-1 data is ambiguous regarding the maximum brightness, which might also be influencing the period determination.  This star was observed to be a SiO maser source in \citet{deguchi2007} and was noted for having a highly unusual radial velocity relative to the Local Standard of Rest, possibly a dynamical effect due to the influence of the Galactic bulge bar \citep{deguchi2010}.  This star was observed in all four wavelengths by \textit{AKARI}/FIS, with equivalent magnitudes in 65, 90, 140 and 160 microns of about 8.2, 9.7, 8.2 and 7.7, respectively.  Some other variability is evident in the lightcurve, possibly due to the nearby variable NSV~12065, this having a period visually between 1 and 2 days.
\item[IRAS 20060-2425]  There are indications of changing brightness of maxima between cycles for this star.  The star has been observed by \textit{IRAS} but is otherwise very poorly observed, although there is a possibility that the nearby AN 958.1936, recorded as variable in \citet{luyten1937}, could have been an observation of it, or that the \textit{STEREO}/HI-1 observations have detected the variability of this star.
\end{description}

\onecolumn

\twocolumn

\subsection{Y Sco: flare-like event or starspot?}
\label{subsec:YSco}

\begin{figure}
\resizebox{8cm}{!}{\includegraphics{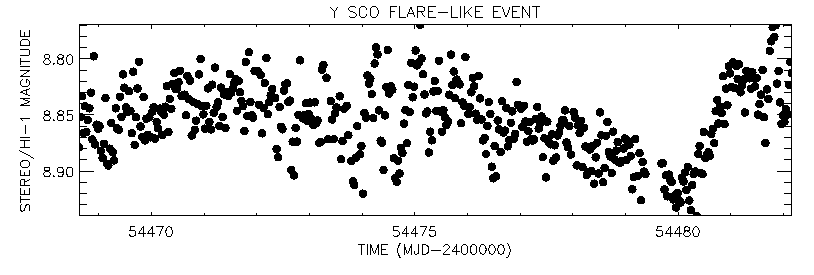}}
\caption{A candidate flare-like event (begins at MJD 2454480) seen in Y Sco by \textit{STEREO}/HI-1B.}
\label{fig16}
\end{figure}

\noindent
A recent search of data from the \textit{CoRoT} mission found at best a  tentative sign of a single flare-like event \citep{lebzelter2011} in only one star.  The larger sample we present here has the difficulty that many potential flares could easily be mistaken for a de-pointing event caused by a micrometeorite impact.  Y Sco, however, displays an event that does not look like any of the known systematics and therefore might be a flare, although the timescale of this feature, being of about 1 day duration, is longer than might be expected (Figure \ref{fig16}).  The feature begins in the first epoch of \textit{STEREO}/HI-1B data, at MJD 2454480 and lasts for one day, during which the brightness increases in a linear fashion by $0.1$~magnitudes.  There is no change in the scatter of the lightcurve during this time, thus it is not due to a micrometeorite hit.  An alternative explanation for this feature might be that it is due to a dark spot on the surface rotating out of view, or perhaps a bright spot rotating into view.  The star is only a few days away from maximum brightness at the time this event occurs.  There is a remark in the GCVS \citep{samus2012} that this star has been irregular since 1972 and although the period observed by \textit{STEREO}/HI-1 of 355~days is very similar to the pre-1972 period of 351.88~days, the interval between the two observed maxima is about 782~days.  If the flare-like event is really due to a starspot, it might also mean that the maximum observed a few days before this was not a full maximum, and the star was still approaching maximum brightness.

Some caution needs to be taken with the variability, as the star actually observed by \textit{STEREO}/HI-1 is NOMAD1 0706-0340442, 2.4~pixels away.  There is very vague short period variability with a period visually close to a day in the lightcurve, with an amplitude of $<$50~mmag that may be from this star and its colours in \citet{zacharias2004} are suggestive of a star of spectral type G or K.  It thus has some potential for rotational variability and flares and spots of its own, although no other indication of such an event is seen in the lightcurve.  If it has a rotational period near 1 day, this would likely exclude it as the source of the flare-like event, since this lasts about as long as one such rotation and could not be due to an event on a part of the surface of this star rotating into or out of view.  There is an unrecorded galaxy with a bright core very close to Y Sco (RA: 247.356 degrees DEC: -19.3638 degrees) but this is probably too faint for any AGN-type effects to be observed by \textit{STEREO}/HI-1.

\subsection{Period Changing Variables}

\noindent
15 of the stars in the sample are recorded in \citet{samus2012} as showing variable periods.  Column 9 of Table \ref{table1} labels these with a 1.  These stars are briefly discussed here for indications of further period changes or other unusual features.  Below this list two other stars showing evidence of period changes are detailed.

\begin{description}
\item[V Tau]  This star has the shortest period of any found in the sample, at 170~days.  This is a very good match to the known period of 168.7~days, thus no sign of period changing is observed.
\item[Z Tau]  This star has the second-longest previously known period of any in the sample, at 466.2~days, whilst the period found here is just outside the typical 4~\verb+%+ difference, at 438~days.  The accuracy of the period determination is unable to exclude a small period change but a significant change is unlikely.
\item[U Ori]  The period found here of 372~days is a good match to the known period of 368.3~days.  The period is probably constant for the duration of the \textit{STEREO}/HI-1 observations.
\item[T CMi]  The period found here of 325~days is a very good match to the known period of 328.3~days.  A change in the shape between cycles is suggested by the phase-folded lightcurve, although a period change might also manifest in this way and not be picked up, thus it is inconclusive for this star whether or not any change has occurred.
\item[S Leo]  The period found here of 197~days is in fair agreement with the known period of 190.16~days.  There is evidence of a small change in the brightness of maxima between cycles.  Two successive maxima are observed with an interval of 186.3~days between them, within the errors of the times of maxima no period change is therefore found.
\item[SS Vir]  The period found here of 377~days is within the typical 4~\verb+%+ away from the known period of 364.14~days.  Two successive maxima are observed with 342.14~days between them, suggesting a period decrease, however.  The phase-folded lightcurve is not smooth and a period change or irregularity could have resulted in a poor period determination.
\item[S Vir]  The \textit{STEREO}/HI-1 period of 356~days is slightly different to the 375.1~days from \citet{samus2012} and the poor phase coverage may be hampering the period determination.  It is therefore uncertain if the period is changing, however the brightness of both maxima and minima might also be changing, which would further complicate the analysis.
\item[S Lib]  The period from \textit{STEREO}/HI-1 of 197~days is in good agreement with the known period of 192.9~days and it seems unlikely that significant period changes have occurred.  The phase-folded lightcurve nevertheless clearly shows that the brightness of maxima and possibly the shape of the lightcurve may differ from cycle to cycle.
\item[Y Sco]  See above discussion in Section \ref{subsec:YSco}.
\item[R Oph]  In spite of the large uncertainty in the period, the \textit{STEREO}/HI-1 period of 302~days is in good agreement with the known period of 306.5~days.  There is an indication of the brightness of minima varying between cycles but no indication of a significant period change.
\item[V1869 Sgr]  The \textit{STEREO}/HI-1 period of 318~days only moderately agrees with the known period of 332~days and the phase coverage is poor as there is less data than normal for this star, likely due to it being so near to the edge of the field of view it may have been missed completely on some orbits.  The analysis is therefore unable to determine if there have been any period changes.
\item[V3876 Sgr]  The period found here of 344~days is in good agreement with the known period of 352~days.  A period change therefore seems unlikely.  There is some indication of a change in maximum brightness between cycles.
\item[RX Sgr]  The period found here of 326~days is in reasonable agreement with the known period of 335.23~days.  An additional complication here is the presence of the known semi-regular variable BH Sgr about 95~arc-seconds away.  Changes in maximum brightness between cycles indicated by the lightcurve are therefore unreliable and it is uncertain what effect this other variable would have on the observations of the period, although with a period from \citet{samus2012} of 100~days and being a much fainter object, the effects of RX Sgr should dominate.
\item[AN Sgr]  The period found here of 325~days agrees moderately well with the known period of 337.56~days.  The phase coverage is mostly concentrated around minimum and there is therefore no evidence of period changing or other significant differences between cycles.
\item[RR Sgr]  The period found here of 335~days is in excellent agreement with the known period of 336.33~days.  The period is therefore unlikely to be significantly changing, however the unfolded lightcurve indicates differences in maximum brightness and possibly shape between cycles.
\end{description}

\begin{figure}
\resizebox{8cm}{!}{\includegraphics{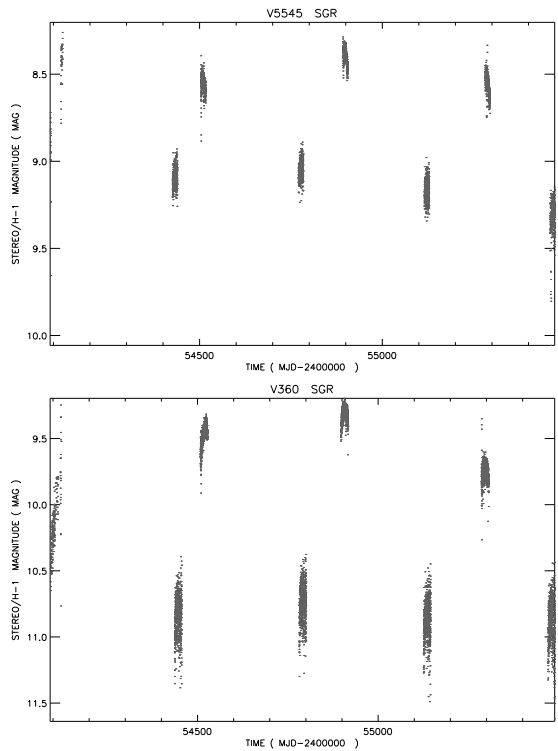}}
\caption{The unfolded \textit{STEREO}/HI-1 lightcurves of V5545 Sgr (upper) and V360 Sgr (lower).  Maxima for V5545 Sgr are observed at (times are MJD-2400000): 54506.99 $\pm 2.5$; 54895.71 $\pm 3.0$ and 55283.38 $\pm 3.0$.  Maxima for V360 Sgr are observed at (times are MJD-2400000): 54506.99 $\pm 2.5$; 54895.71 $\pm 3.0$ and 55283.38 $\pm 3.0$.}
\label{fig17}
\end{figure}

Two other stars show evidence of a different period from that given in \citet{samus2012}.  Both are classified as semi-regular variables rather than Miras.  Their unfolded lightcurves are shown in Figure \ref{fig17}.  Unfortunately, both are too poorly-observed from the ground for the observations to be confirmed - no data for them exists in the online archives of the AAVSO, the BAA VSS or the AFOEV.  Their position on the Ecliptic Plane makes long-term monitoring difficult and without long-term monitoring it is impossible to ascertain whether the periods are actively changing or meandering about a more regular value \citep{zijlstra2002}.  The two stars are
\begin{description}
\item[V5545 Sgr]  Although the period found here of 368~days is in reasonable agreement with the known period of 377~days, the three consecutive times of maximum brightness observed indicate a period of about 388~days (Figure \ref{fig17}, upper plot).  A small period change may therefore have occurred.
\item[V360 Sgr]  The period found here of 367~days is, at best, a very rough approximation of a harmonic of the known period of 165~days.  The three maxima observed are not consistent with either period, being separated by 383.47~days and 389.39~days (Figure \ref{fig17}, lower plot).  A significant period change may therefore have occurred, or the period may be irregular.
\end{description}

\section{\uppercase{Discussion}}
\label{sec:discussion}

\noindent
The arrangement of two almost identical cameras on almost identical satellites in different heliocentric orbits permits the observation of periodicity on timescales not observable from the Earth or from a single satellite alone.    The unusual bandpass of the \textit{STEREO}/HI-1 instruments, in particular the throughput in the infra-red at about 950~nm (Figure \ref{fig1}), allows for the observation of stars of very late spectral type and also for stars that are obscured by circumstellar dust shells.  Together, this provides homogenous observations of Mira variables with improved phase coverage and has potentially detected some new objects of this type or the related semi-regular class.

Even though some objects have either no \textit{R} magnitude listed in \citet{zacharias2004} or are fainter than the cutoff of 12th magnitude used to select objects to be observed by \textit{STEREO}/HI-1, the large pixel scale of 70 arc-seconds nevertheless allows them to have an indirect influence on the lightcurves of very nearby stars that are in the \textit{STEREO}/HI-1 database (Figure \ref{fig4}).  Although in some regions, the Galactic centre especially, it is difficult or impossible to ascertain the likely source of variability, the very large amplitudes and very long periods of Miras and semi-regular variables is sufficiently distinctive that in some cases the variability is reliably observed even in a densely populated field of view.  In the case of new candidate variables, it further complicates the matter of classification, which is essentially impossible from \textit{STEREO}/HI-1 data alone, although it is likely most are of the semi-regular class.

The majority of the sample of 85 stars are known Mira variables, with a small minority classified as semi-regular variables and a single variable of the Orion type.  7 are listed as variable in either the GCVS or NSV \citet{samus2012} but without a classification and 6 have not been previously observed to be variable.  For 19 stars, the period we present here is the first determination of a period.  There is a reasonable agreement in the periods previously known and those found by an analysis of the \textit{STEREO}/HI-1 photometry (Figure \ref{fig9}), although a harmonic is occasionally found and also some of the stars are known to change their periods.  The accuracy of the period determination is insufficient to detect small changes in period of a few days but larger changes of a couple of weeks are potentially observable (Figure \ref{fig8}).  The new candidate and unclassified variables are individually discussed, as are those known or suspected of period changing.  One star showing a particularly unusual feature is discussed: Y Sco, which shows a candidate flare-like event, although a starspot might also be an alternative explanation (Figure \ref{fig16}).

Of the newly-discovered variables, two (IRC -20507 and 2MASS J19291709-2034504) have exceptionally large values of $B2-I$ (Figure \ref{fig12}).  This may be suggestive of the presence of a circumstellar envelope and these stars might be therefore potential maser sources, indeed IRC -20507 has already been recorded as such \citep{Hekkert1991OHmaser1612}.  Follow-up observations of these stars would contribute to the understanding of the evolution of stars during the AGB stage.  There are features of some stars of this type that cannot be explained by current models, such as an excess of ammonia \citep{menten2010} and sub-solar values of $^{16}$O/$^{17}$O and $^{16}$O/$^{18}$O \citep{decin2010}, and having more examples to study would be very useful.  We encourage interested readers to make follow-up observations of the new candidate variables, in order to confirm their nature and periodicity.  To assist in this, we provide dates when the new candidate variables will be observed by \textit{STEREO}/HI-1 so that observations may be conducted simultaneously (Table \ref{table2}).  It is hoped that five years' worth of \textit{STEREO}/HI-1 data will soon be made available for all stars in the field of view with listed \textit{R} magnitudes of 12 or greater in the NOMAD1 catalogue \citep{zacharias2004}, although a date has not yet been fixed.  In due course, more data will be gathered by \textit{STEREO}/HI but it is not known when, or even if, this will be made available on a large scale.

A complete lack of objects with periods between 200 and 300~days is observed in the sample (Figure \ref{fig10}, right).  There are a small number found with periods that are over 400~days thus it cannot be completely excluded that one or more of those may in fact be harmonics with a genuine period in this range.  Nevertheless, for a sample of this size, this is a significant feature and the only likely explanation is that it is the result of a selection effect.  In the earliest stage of the analysis, many thousands of lightcurves were visually examined in a search for long period variability, however it is not easy to distinguish artificial effects from genuine variability, especially for stars with potentially very erratic variability in terms of the magnitude of maxima, minima, shape and even period.  This may have caused genuine variables with periodicity in this range to be too unconvincing to have been recorded as a likely variable.  A similar selection effect might also have led to a preference for finding variables with periods near 1 year or close to the orbital periods of the two \textit{STEREO} satellites as these produce a more recognisable pattern in the unfolded lightcurves.  It is therefore likely that more Miras and semi-regular variables remain undiscovered in the data.

A selection effect may also be responsible for the prevalence of stars in our sample with large $B2-I$ values from \citet{monet2003}, shown on the right in Figure \ref{fig12}.  The search for the sources of variability was focused on very red objects, although other colours are also given in \citet{zacharias2004}, in particular the $J$, $H$ and $K$ colours from \citet{skrutskie2006}.  This prevalence could also be partly due to the sensitivity of the \textit{STEREO}/HI-1 imagers at about 950~nm (Figure \ref{fig1}), making it easier to recover variability from stars bright at this wavelength.

The photometry gathered by \textit{STEREO}/HI-1 on both long and short timescales of bright stars on the Ecliptic Plane is a valuable resource for the observation and monitoring of Miras and semi-regular variables.  Some variables of this type have periods close to a year, or a fraction thereof, and the full range of their behaviour cannot be monitored from Earth or Earth orbit and thus the homogenous observations of \textit{STEREO}/HI-1 provide a unique window to advance the study of these poorly-understood objects.

\section*{\uppercase{Acknowledgements}}
\label{sec:acks}

\noindent The Heliospheric Imager (HI) instrument was developed by a collaboration that included the Rutherford Appleton Laboratory and the University of Birmingham, both in the United Kingdom, and the Centre Spatial de Li\'ege (CSL), Belgium, and the US Naval Research Laboratory (NRL), Washington DC, USA.  The \textit{STEREO}/SECCHI project is an international consortium of the Naval Research Laboratory (USA), Lockheed Martin Solar and Astrophysics Lab (USA), NASA Goddard Space Flight Center (USA), Rutherford Appleton Laboratory (UK), University of Birmingham (UK), Max-Planck-Institut f\"{u}r Sonnensystemforschung (Germany), Centre Spatial de Li\'ege (Belgium), Institut d'Optique Th\'eorique et Appliqu\'ee (France) and Institut d'Astrophysique Spatiale (France).  This research has made use of the \textsc{Simbad} database, operated at CDS, Strasbourg, France.  This research has made use of NASA's Astrophysics Data System.  This research has made use of the statistical analysis package \textsc{R} \citep{rproject}.  This publication makes use of data products from the Two Micron All Sky Survey, which is a joint project of the University of Massachusetts and the Infrared Processing and Analysis Center/California Institute of Technology, funded by the National Aeronautics and Space Administration and the National Science Foundation.  This research has made use of version 2.31 \textsc{Peranso} light curve and period analysis software, maintained at CBA, Belgium Observatory http://www.cbabelgium.com.  This research is based on observations with AKARI, a JAXA project with the participation of ESA.  This research is funded by the Science and Technology Facilities Council (STFC). KTW acknowledges support from a STFC studentship.  This research was funded by the Austrian Science Fund (FWF): P21988-N16 and AP2300621.  The authors wish to thank Dr. Luca Fossati of the Open University for useful conversations.

\appendix

\section{Phase-folded lightcurves}
\label{sec:appendix}

\begin{figure*}
\resizebox{17cm}{!}{\includegraphics{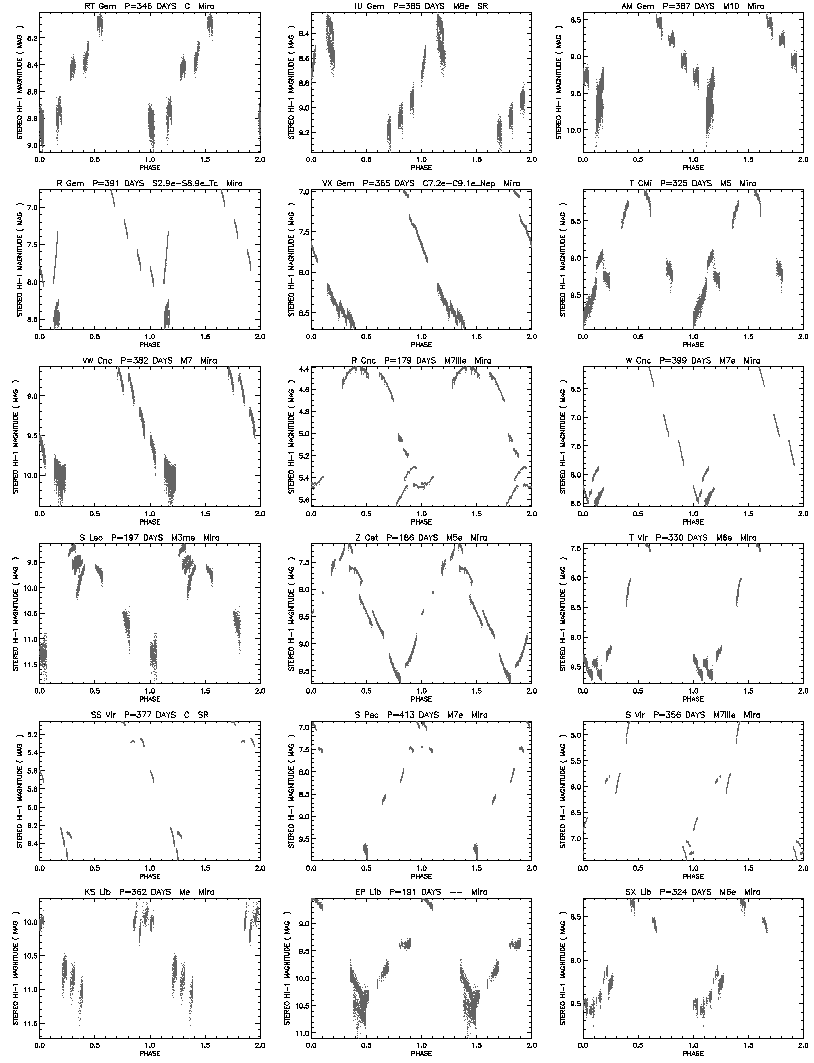}}
\caption{Phase-folded lightcurves of stars showing long period variability in \textit{STEREO}/HI-1.  2 phases of the periods indicated are shown.}
\label{figA1}
\end{figure*}

\begin{figure*}
\resizebox{17cm}{!}{\includegraphics{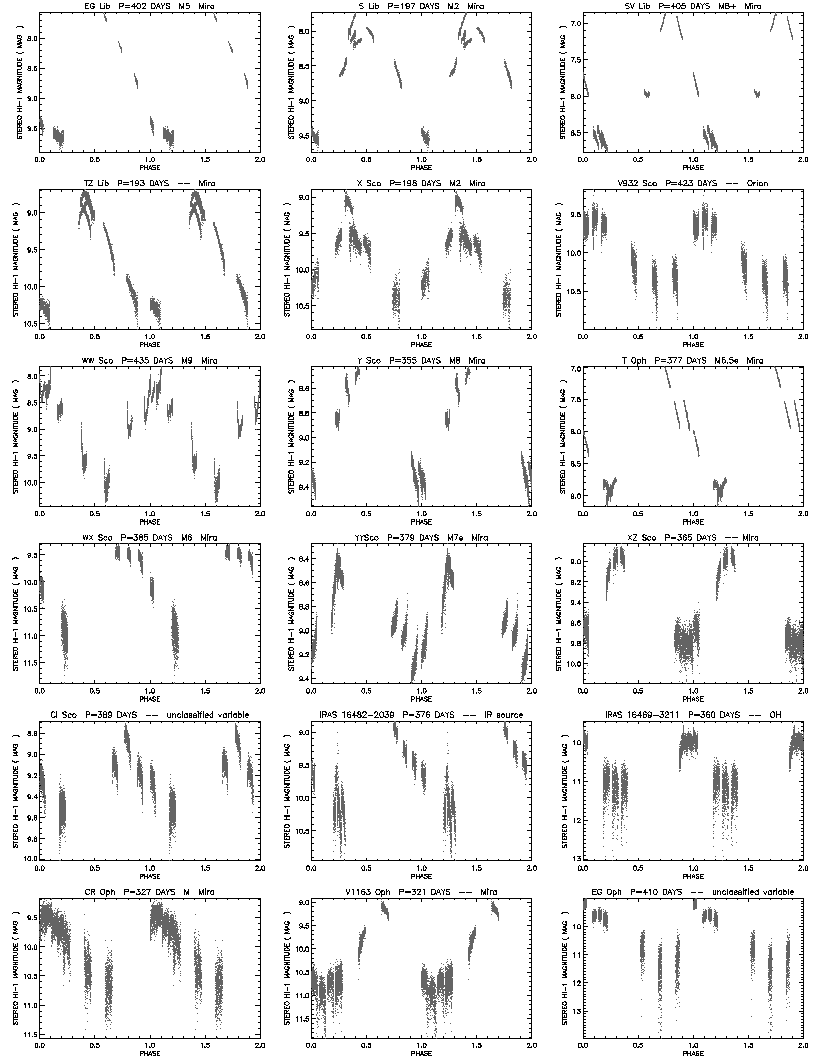}}
\caption{Phase-folded lightcurves of stars showing long period variability in \textit{STEREO}/HI-1.  2 phases of the periods indicated are shown.}
\label{figA2}
\end{figure*}

\begin{figure*}
\resizebox{17cm}{!}{\includegraphics{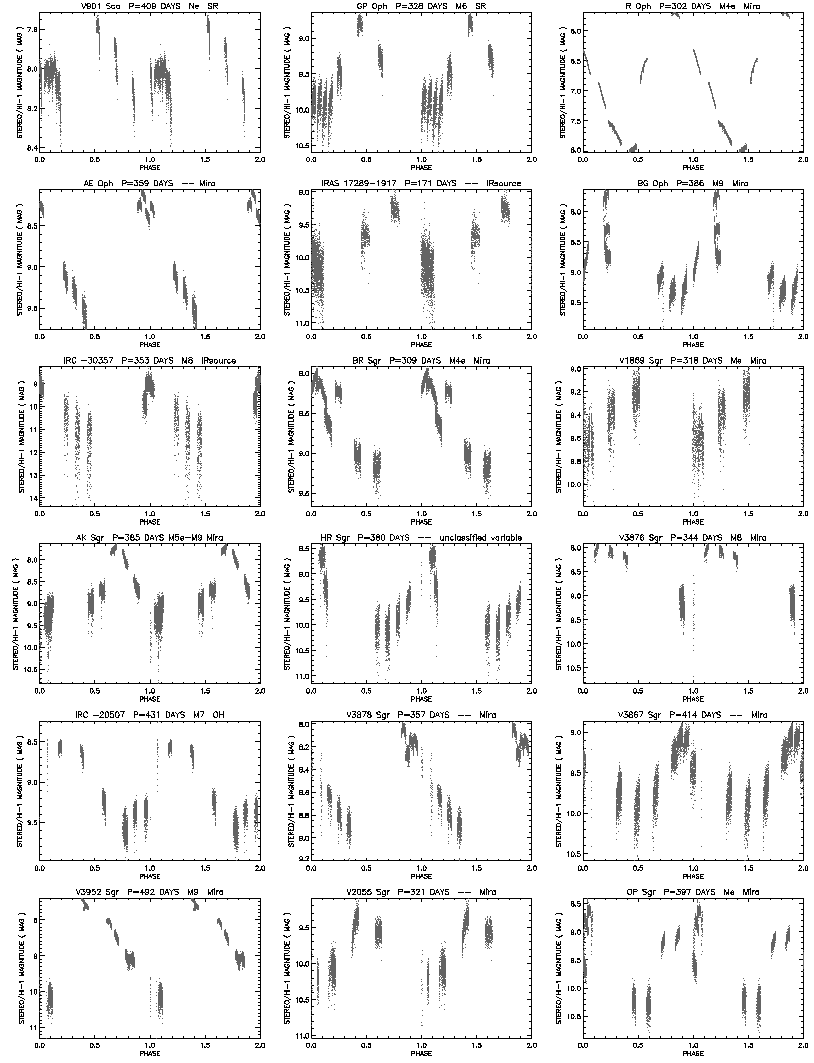}}
\caption{Phase-folded lightcurves of stars showing long period variability in \textit{STEREO}/HI-1.  2 phases of the periods indicated are shown.}
\label{figA3}
\end{figure*}

\begin{figure*}
\resizebox{17cm}{!}{\includegraphics{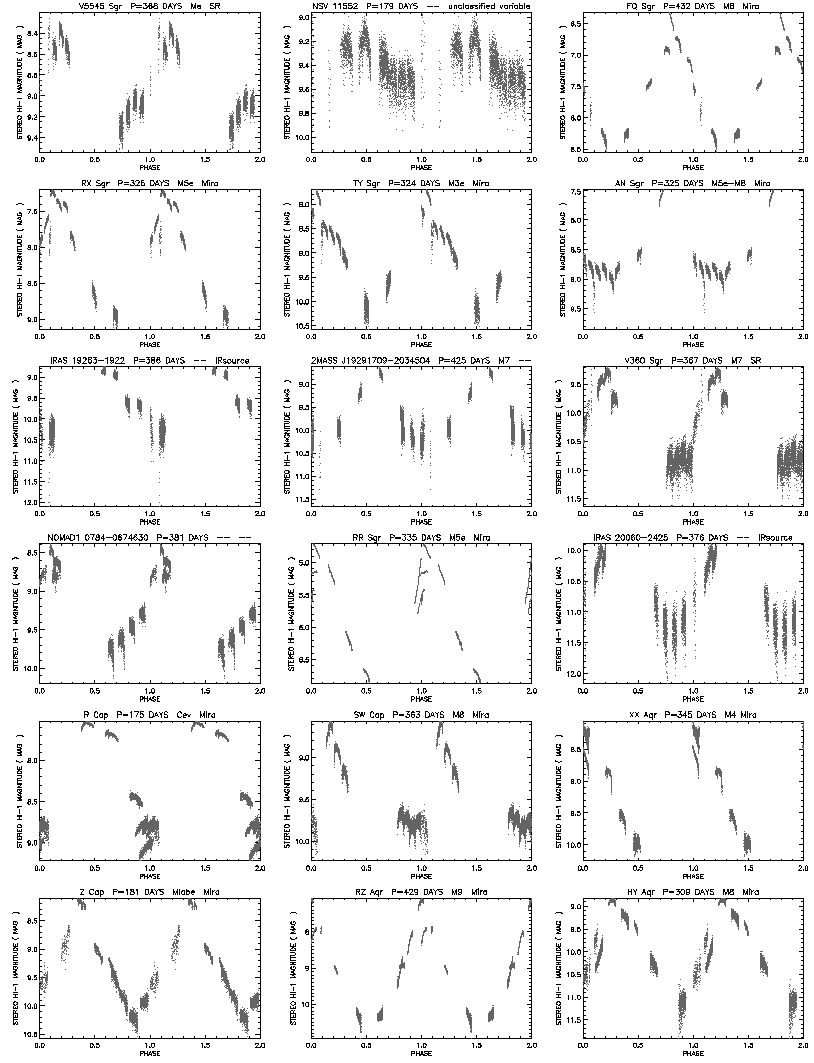}}
\caption{Phase-folded lightcurves of stars showing long period variability in \textit{STEREO}/HI-1.  2 phases of the periods indicated are shown.}
\label{figA4}
\end{figure*}

\begin{figure*}
\resizebox{17cm}{!}{\includegraphics{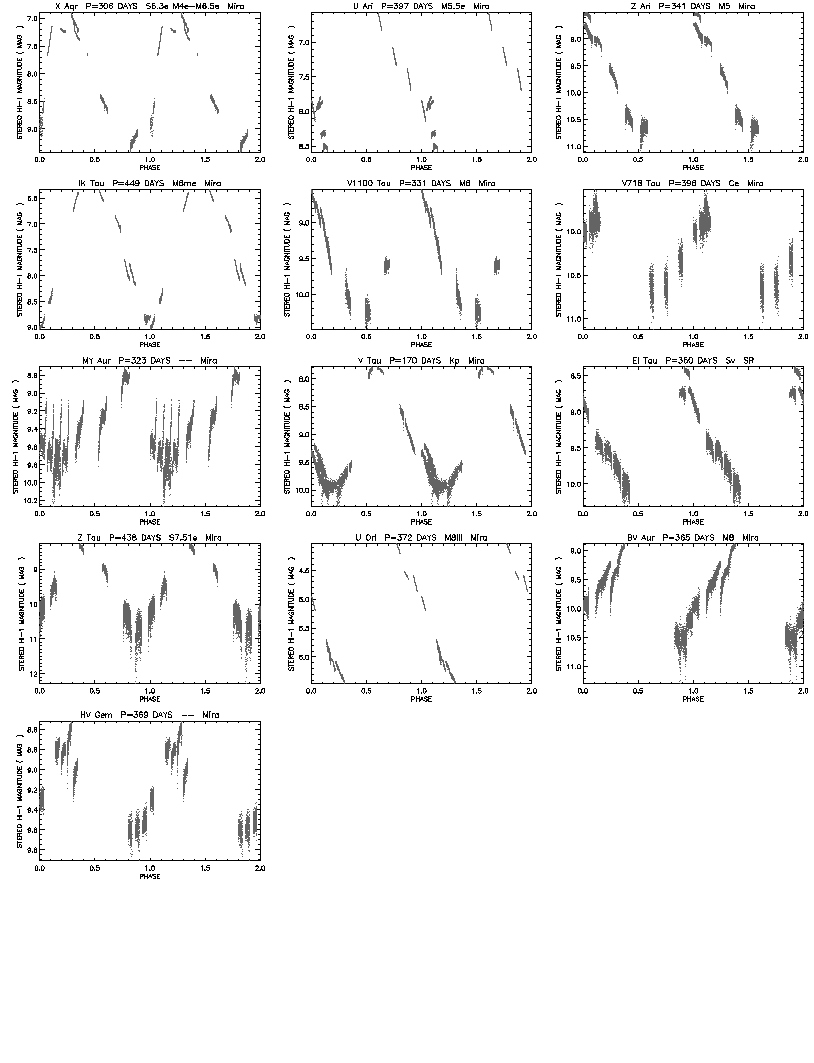}}
\caption{Phase-folded lightcurves of stars showing long period variability in \textit{STEREO}/HI-1.  2 phases of the periods indicated are shown.}
\label{figA5}
\end{figure*}

\noindent
The phase-folded lightcurves for all 85 stars in the sample are shown in Figures \ref{figA1}, \ref{figA2}, \ref{figA3}, \ref{figA4} and \ref{figA5}.  In each case, the period folded on is shown in the title, along with some other information on the star for ease of reference.

\onecolumn

\twocolumn

\end{document}